\documentclass[12pt,a4paper]{article}

\usepackage[utf8]{inputenc}
\usepackage{fullpage}
\usepackage{hyperref}
\usepackage{amssymb}
\usepackage{amsthm}
\usepackage{graphicx}
\usepackage[outdir=./]{epstopdf}
\usepackage{color}
\usepackage{enumitem}
\usepackage{empheq}
\usepackage{cite}
\usepackage{slashed}
\numberwithin{equation}{section}
\usepackage{tikz}
\usetikzlibrary{arrows}
\usetikzlibrary{decorations.markings}
\usetikzlibrary{shapes}
\tikzset{
	label/.style={
		postaction={
			decorate,
			decoration={
				markings,
				mark=at position 0.5 with {\arrow{stealth}},
				mark=at position 0.5 with \node #1;
			}
		}
	},
	contourlabel/.style={
		postaction={
			decorate,
			decoration={
				markings,
				mark=at position 0.5 with {\fill (2pt,0)--(-2pt,2.31pt)--(-2pt,-2.31pt)--cycle;},
				mark=at position 0.5 with \node #1;
			}
		}
	},
	gluon/.style={
		decorate,draw=black,decoration={coil,aspect=0.8,amplitude=3pt, segment length=5pt}
	},	
	softgluon/.style={
		decorate,draw=red,decoration={coil,aspect=1.2,amplitude=2pt,segment length=7pt}
	},
	quark/.style={
		postaction={decorate,decoration={markings,mark=at position 0.5 with 
				{\fill (2pt,0)--(-2pt,2.31pt)--(-2pt,-2.31pt)--cycle;}
		}}
	},
	boson/.style={
		decorate,draw=black,decoration={snake,amplitude=3pt, segment length=6pt}
	},
	higgs/.style={
		dashed
	},
	zigzag/.style={
		decorate,draw=black,decoration={zigzag,amplitude=3pt, segment length=6pt}
	},
	correspondance/.style={
		blue,->, >=latex', shorten >=1pt, thick
	}		
}
\usepackage{pgfplots}
\ifdim\overfullrule>0pt 
\usepackage{environ}

\NewEnviron{tikzpicture}{%
	\begin{pgfpicture}
		\pgfpathrectanglecorners{\pgfpointorigin}{\pgfpoint{3cm}{3cm}}%
		\pgfusepath{stroke}\end{pgfpicture}%
}
\fi

\usepackage{cleveref}
\crefname{section}{Section}{Sections}
\crefname{appendix}{Appendix}{Appendices}

\newcommand{\WT}[1]{ \widetilde W_{#1}(0,\infty)}

\begin{document}

\titlepage

\begin{flushright}
	MS-TP-20-33
\end{flushright}

\vspace*{1.2cm}

\begin{center}
	{\Large \bf Asymptotic dynamics on the worldline for spinning particles}
	
	\vspace*{1cm} \textsc { Domenico Bonocore} \\
	
	\vspace*{1cm}

Institut f\"{u}r Theoretische Physik, Westf\"{a}lische
	Wilhelms-Universit\"{a}t M\"{u}nster, Wilhelm-Klemm-Stra\ss e 9,
	D-48149 M\"{u}nster, Germany\\

\end{center}

\vspace*{7mm}

\begin{abstract}
	\noindent 
	
	There has been a renewed interest in the description of dressed asymptotic states
	 \`a la Faddeev-Kulish.
	In this regard, 
	a worldline representation for asymptotic states dressed by radiation at subleading power
	in the soft expansion,
	 known as the Generalized Wilson Line (GWL) in the literature, has been available for some time,
	 and it recently 
	  found applications in the derivation of factorization theorems for scattering 
	  processes of phenomenological relevance. 
	In this paper we revisit the derivation of the GWL in the light 
	of the well-known supersymmetric wordline formalism for the relativistic spinning
	particle. In particular, we discuss the importance of
	 wordline supersymmetry to understand the contribution
	 of the soft background field to the
	 asymptotic dynamics. 
	We also provide a derivation of the GWL for the gluon case, 
	which was not previously available in the literature, thus extending the 
	exponentiation of next-to-soft gauge boson corrections to Yang-Mills theory.
	  Finally, we 
	comment about 
	possible applications in the current research about asymptotic states 
	in scattering amplitudes for gauge and gravity theories
	and their classical limit.

\end{abstract}


\newpage

\begingroup
\hypersetup{hidelinks}
\tableofcontents
\endgroup

\newpage

\section{Introduction}

The use of first-quantized techniques in quantum field theories has a long history. 
The origin can be traced back 
to the early days of quantum electrodynamics, when Schwinger \cite{Schwinger:1951nm}  proposed to use
the distribution identity
\begin{align}
	\lim_{\epsilon\to 0^+}\frac{i}{H+i\epsilon}=\lim_{\epsilon\to 0^+}\int_0^{\infty}dT\, e^{i(H+i\epsilon) T}
	\label{schwinger}
\end{align}
to interpret the inverse propagator of a scalar field  
as the matrix elements of a
Hamiltonian $H(\hat x_{\mu},\hat p_{\mu})$
governing the evolution in proper time $T$ of a first-quantized model
with canonical variables $ \hat x_{\mu}$ and $ \hat p_{\mu}$.
By writing a path integral representation for the matrix elements
of the evolution
operator $e^{iHT}$,
the dynamics of the relativistic particle
then is described by 
a classical action for the fields $x^{\mu}(t)$ and $p^{\mu}(t)$
leaving in a one dimensional space of length~$T$. 
The
worldline formalism \cite{Schubert:2001he} is based on the observation that such actions can be derived
from first principles,
starting 
from the constrained quantization of the relativistic particle rather than 
the field theory propagator.

A major obstacle for the program is given by the presence of matrices in the exponent, which occur e.g. for spinning or colored particles. 
%
%
This in general requires either a path ordering prescription or the introduction of Grassman variables $\psi_{\mu}(t)$. With the latter option, the action for 
a free massless particle of spin $N/2$ reads \cite{Barducci:1976qu, Brink:1976sz, Brink:1976uf, Gershun:1979fb}
\begin{align}
	S=~\int dt\,\left( p_{\mu}\dot{x}^{\mu}
	+ \frac{i}{2}\psi_i^{\mu}\dot{ \psi^i_{\mu}}
	-\frac{1}{2}ep_{\mu}p^{\mu}
	-i\chi_i\psi^{i}_{\mu}p^{\mu}-\frac{i}{2}a_{ij}\psi_i^{\mu} \psi^j_{\mu}\right)~,
	\label{susy}
\end{align}
with $i=1,...,N$.
The symmetry structure of this action is quite rich, since it involves  reparametrization invariance, $N$-extended local supersymmetry and $O(N)$ gauge invariance, with the corresponding gauge fields $e(t)$, $\chi_i(t)$ and $a_{ij}(t)$. As is evident from \cref{susy}, these fields play the role of Lagrange multipliers which are typically gauged away for practical calculations.   

Most of the early work (see e.g. \cite{Howe:1988ft, Howe:1989vn, Bergshoeff:1989uj, Gitman:1994wc}) focused on  quantization issues of the supersymmetric model and on attempts to formulate it on curved spacetime. However, an application showing the practical  advantages of this formulation for actual calculations remained somewhat elusive. In this regard, a major advance was put forward by  Strassler\cite{Strassler:1992zr}. The idea stems from the fact that a one-loop effective action in the background of some gauge boson field can be described as a dressed propagator whose extrema have been closed to form a loop. 
Using a worldline representation for such a propagator, Strassler verified that upon solving the path integral order by order in the coupling constant $g$, one is left at order $g^N$ with the amplitude for $N$ external gauge bosons in terms of integrals over Feynman parameters only, thus bypassing the construction of the Feynman amplitude via standard Feynman rules. In this way he recovered the Bern-Kosower rules \cite{Bern:1991aq} previously derived from string theory insights.

Motivated by this success, most of the  applications of the worldline approach have aimed at the computation of effective actions (see e.g. \cite{Schmidt:1993rk, Schmidt:1994zj}). On the other hand, the worldline representation of an \emph{open} dressed propagator has received less attention, with intermittent progress  \cite{DiVecchia:1979gx,Fradkin:1991ci,Pierri:1990rp, Marnelius:1993ba, vanHolten:1995ds, Reuter:1996zm, Alexandrou:1998ia, Dai:2008bh, Ahmadiniaz:2015xoa, Bhattacharya:2017wlw},
although it
has a richer structure that might reveal new
methods for the efficient calculation of scattering amplitudes.\footnote{For recent progress in this direction see \cite{Ahmadiniaz:2020wlm, Corradini:2020prz}.}

A situation which is somewhat intermediate between the open and the closed dressed propagators is given by \emph{asymptotic} dressed propagators. This is what was considered  
 by Laenen, White and Stavenga\cite{Laenen:2008gt}, who combined the aferomentioned body of work
 and the long-standing  problem of infrared exponentiation beyond the leading power in the soft expansion (there dubbed next-to-eikonal) \cite{Gatheral:1983cz, Frenkel:1984pz, Gardi:2010rn,  Gardi:2013ita}. 
  Such exponentiation emerges neatly in this picture by solving the path integral representation of the dressed propagator order by order in the soft expansion (but to all-orders in the coupling constant). More specifically, in the abelian case it corresponds to the familiar exponentiation of connected diagrams, while for non-abelian theories one has to invoke the so-called ``replica trick'' from statistical physics. This bypasses highly non-trivial combinatorics that one should use in a purely diagrammatic approach \cite{Laenen:2010uz}, showing the power of the worldline formulation.  
 The approach has also been implemented for soft gravitons in \cite{White:2011yy}.

Building on this work, and motivated by the rising interest in the field of next-to-leading power (NLP)  corrections to the soft and collinear limits both in phenomenology \cite{Laenen:2010kp, Moch:2009hr, Bonocore:2014wua, Bonocore:2015esa, Moult:2018jjd, Moult:2019mog,  Moult:2019vou, Ebert:2018lzn, Anastasiou:2015vya, Bahjat-Abbas:2018hpv, vanBeekveld:2019cks, vanBeekveld:2019prq, DelDuca:2017twk, Beneke:2017ztn, Beneke:2018rbh, Beneke:2018gvs, Beneke:2019oqx, Gervais:2017yxv, Laenen:2020nrt} and in more formal contexts \cite{Strominger:2013jfa, Cachazo:2014fwa, Casali:2014xpa, Bern:2014oka, Larkoski:2014bxa,He:2014bga, Larkoski:2014hta,Afkhami-Jeddi:2014fia, Adamo:2014yya, He:2014laa, Klose:2015xoa, Brandhuber:2015vhm, Sen:2017nim, Laddha:2017ygw, Sahoo:2018lxl,   Laddha:2018myi, Sahoo:2020ryf}, the worldline description of \cite{Laenen:2008gt} has proved to be a valuable tool to derive factorization theorems at NLP \cite{Bonocore:2016awd, Bahjat-Abbas:2019fqa}. 
The asymptotic dressed propagator defined in this way at NLP has been dubbed Generalized Wilson Line (GWL), and 
 it is defined for a semi-infinite straight line 
  starting from the origin in the direction $n^{\mu}$ as
\begin{align}
\WT{n}
&={\cal P} \exp \left[ \, g \! \int \frac{d^d k}{(2 \pi)^d} \tilde A_\mu(k)
\left( - \frac{n^\mu}{n\cdot k} + \frac{k^\mu}{2 n \cdot k} - 
k^2 \frac{n^\mu}{2 (n \cdot k)^2} - \frac{{\rm i} k_\nu J^{\nu\mu}}
{n \cdot k} \right) \right. \notag \\
& \,\, \, + \left. \int \frac{d^d k}{(2 \pi)^d} \int \frac{d^d l}{(2 \pi)^d}
\tilde A_\mu(k) \tilde A_\nu(l) \left(\frac{\eta^{\mu \nu}}{2 n \cdot (k + l)} - 
\frac{n^\nu l^\mu n \cdot k + n^\mu k^\nu n \cdot l}{2 (n \cdot l)(n \cdot k) 
	\left[n \cdot (k + l) \right]} \right.\right. \notag \\
& \,\,\,  + \left. \left. \frac{(k \cdot l) n^\mu n^\nu}{2 (n \cdot l)(n \cdot k)
	\left[n \cdot (k + l) \right]} 
- \frac{{\rm i} J^{\mu\nu}}{n \cdot (k + l)}
\right) \right] \, ,
\label{wilson}
\end{align}
where ${\cal P}$ is the path-ordering symbol and $\tilde A_{\mu}(k)=\tilde A_{\mu}^a(k)T^a$ is the Fourier transform of the non-abelian gauge field corresponding to a soft gluon emission of momentum $k^{\mu}$.  The term of order $k^{-1}$ in the first line of \cref{wilson} corresponds to the usual Wilson line for a semi-infinite straight path, while the remaining terms correspond to NLP corrections.  
Two of these subleading terms contain the total 
angular momentum $J^{\mu\nu}$, which 
for an emitting particle in the representation specified by the indices $i$ and $j$
is given by the sum 
 of the 
Lorentz generators for spin  $(S^{\mu\nu})^{ij}$ and angular momentum $L^{\mu\nu}\delta^{ij}$.
\footnote{The angular momentum $L^{\mu\nu}$ was not included in the definition provided in 
\cite{Laenen:2008gt, Bonocore:2016awd, Bahjat-Abbas:2019fqa}, since the corresponding 
\emph{internal} emissions from Low's theorem give rise to contributions
that do not
exponentiate to all-orders, unlike the other terms which are due to \emph{external} emissions. However, as observed in \cite{Bahjat-Abbas:2019fqa},
the separation between internal and external emissions is not gauge invariant. Hence, 
in order for the Generalized Wilson Line to transform better under gauge tranformations,
 it might be convenient to include the angular momentum in the definition. \label{footnote:gauge}
}

The validity of \cref{wilson} for generic spin has been motivated  by the result one would expect for the closed propagator of the one-loop effective action \cite{Strassler:1992zr, Reuter:1996zm}, where the coupling between the spin and the background field is contained exclusively in the chromo-magnetic interaction $S_{\mu\nu}F^{\mu\nu}$. In fact, after replacing the  external lines of a scattering amplitude with the straight and semi-infinite GWLs, these close at infinity at the cross-section level, by unitarity. Therefore, one should expect the 
one-loop effective action description to be equivalent when the formalism is implemented in physical processes. Moreover, this picture has been corroborated for tree-level amplitudes with a single soft emission by the recently discovered next-to-soft theorems \cite{Cachazo:2014fwa}, and a comparison has been provided in \cite{White:2014qia}.

However, one would like a less heuristic argument that could put
\cref{wilson} on a solid ground for a generic scattering amplitude. In fact,
   a derivation of \cref{wilson} from the worldline representation of an open propagator of generic spin is still missing. 
 Even for the Dirac case, which was analyzed in \cite{Laenen:2008gt}, some issues still remain to be clarified.
  For instance, in that case \cref{wilson} has been obtained by first decomposing the dressed Dirac propagator in terms of the covariant derivative $D_{\mu}=\partial_{\mu}-A_{\mu}$ as\footnote{Unlike \cite{Laenen:2008gt}, we use the $(+,-,-,-)$ metric throughout.} 
\begin{align}
	 \frac{1}{i\slashed{D}-m}=\frac{i\slashed D+m}{-\slashed D^2-m^2}
	 =\frac{i\slashed D+m}{-D^2-m^2+S^{\mu\nu}F_{\mu\nu}}~,
	\label{dressedDirac}
\end{align} 
with $S^{\mu\nu}=\frac{i}{4}[\gamma^{\mu},\gamma^{\nu}]$, and then writing a worldline representation for the denominator only. In fact, the numerator does not contribute for the one-loop effective action\cite{Strassler:1992zr}. On the other hand, as remarked in \cite{Laenen:2008gt}, for an asymptotic dressed propagator the numerator \emph{does} contribute but it is supposed to cancel \`a la LSZ with the numerator of the \emph{free} inverse propagator. However, this is obviously correct only if the gauge field contribution in the numerator vanishes. What is the mechanism behind that?
 
Things get even more subtle for spin-one (which was not discussed in \cite{Laenen:2008gt}), where in the massless case one has to deal with the gauge dependence of the emitting particle. In Feynman gauge, the numerator of the dressed propagator is unity and thus
one expects the  argument to mimic what done in the case of the one-loop effective action. However, a more precise derivation would be desirable. Moreover, one would like to extend the validity of \cref{wilson} also to massive vector particles, and possibly to higher spin.     

Quite generally, what is missing is a clear relation between the supersymmetric formulation of the relativistic spinning particle (i.e. the equivalent of \cref{susy} with a background field) and the generalized Wilson line
of \cref{wilson}. This is the main goal of this paper and in fact it will turn out that a clear relation between the two descriptions will answer the previous questions, putting the derivation of the GWL on a firm basis also for spinning particles.

As we have remarked, the need for a clear derivation of the GWL for the case 
of spinning particle is mainly 
of phenomenological origin, since QCD  
scattering processes involve quarks and gluons in the initial and final states. 
In this regard, the GWL turned out to be useful
in the recent attempts to extend the traditional soft-gluon resummation program to NLP. 
In particular, 
the GWL
 has been already implemented in the Leading-Logarithmic NLP soft-gluon resummation
 for the inclusive production
 of color-singlet states, such as 
 Higgs production via gluon-fusion \cite{Bahjat-Abbas:2019fqa},
 although a formal proof from first principles for the spin-one case has been lacking.
 This last point makes a derivation of the spinning GWL from first principles even more desirable. 
 
However, the GWL have a much broader scope, which offers other  
 motivations for the present study.
 One arises from a revived interest in asymptotic states \`a la Faddeev-Kulish
 \cite{Kulish:1970ut, Catani:1985xt, DelDuca:1989jt, Kapec:2017tkm, Strominger:2017zoo, Pasterski:2017kqt, Carney:2018ygh,
 	Gonzo:2019fai, Choi:2019rlz, Pate:2019lpp , Law:2020tsg, Casali:2020vuy, Narayanan:2020amh   }.
  Most of these methods revolve around the existence of an asymptotic Hamiltonian governing
  the evolution of the asymptotic states. In particular, reference
   \cite{Hannesdottir:2019rqq, Hannesdottir:2019opa}, building
  on insights from soft-collinear effective theories,
  provides a systematic calculation of such Hamiltonians that makes the S-matrix infrared finite.
  The use of GWLs in this context would easily extend such description beyond the leading power in the soft expansion.
  Indeed, this could provide more efficient definitions of infrared-finite S-matrices,
   allowing to move effects from the asymptotic to the S-matrix. To this end, it is therefore desirable to have a firm derivation of the GWL also for spinning particles.  

More generally, 
there has been recently a great deal of interest in the classical limits 
of scattering amplitudes \cite{Kosower:2018adc, Maybee:2019jus,  delaCruz:2020bbn}, mainly motivated by 
the growing interest in precision
calculations in gravitational physics.
In particular,
there is evidence that the
 high-energy-limit and the corresponding eikonal approximation
are key to extract the classical limit \cite{Amati:1987wq, Damour:2017zjx, KoemansCollado:2019ggb}.
In this context, the semiclassical picture of the GWL extends
the eikonal approximation to subleading power, and thus 
 provides a new tool to study the classical limit of 
 scattering amplitudes. Work in this direction has been done in the scalar case in  \cite{Melville:2013qca, Luna:2016idw}. Besides, 
the GWL might provide an efficient way to extend the classical limit of 
soft theorems to subleading power 
\cite{Sahoo:2018lxl, Laddha:2018myi, A:2020lub}. 
Therefore, given the importance of spin in gravitational physics \cite{Levi:2015ixa,  Vines:2016qwa, Guevara:2018wpp, Bern:2020buy, Antonelli:2020aeb,  Matas:2020wab}, it would be desirable to extend the derivation of the GWL 
to particles of arbitrary spin.


The structure of the paper is the following. We begin in \cref{sec:scalar}
by revisiting the scalar case originally presented in \cite{Laenen:2008gt}. 
The goal here is to highlight the relation with the worldline formalism
 and its symmetries, stressing the 
distinctive features that arise for asymptotic propagators dressed by soft radiation.
Then, in \cref{dirac} we will move to the Dirac case, where the supersymmetry of the model will 
allow us to 
write the dressed propagator in terms of conserved charges and subsequently to observe that
the soft field in the numerator does not contribute in the asymptotic limit. 
Finally, in \cref{sec:gluons} we will discuss the spin-one case, where we will 
first justify the definition of the GWL for gluons without wordline fermions. Then
we will discuss the corresponding supersymmetric model, 
paving the way for a generalization to particles of higher spin.  
We conclude in \cref{sec:discussion} with a short discussion. 


\section{Spin zero}
\label{sec:scalar}

%
Although the generalized Wilson line for a scalar particle has been already discussed in \cite{Laenen:2008gt}, it is useful to revisit the derivation in a different approach, i.e. starting from the constrained quantization of the relativistic particle, which is more standard in the worldline literature, highlighting  the distinctive features that appear in the case of an asymptotic propagator dressed by soft radiation.

\subsection{Dressed propagators and conserved charges}
We start from the well-known\footnote{See e.g. \cite{Banados:2016zim} for a pedagogical review.}
phase space action for a free relativistic scalar particle
\begin{align}
	S=\int dt \left(p\cdot \dot{x} -e\frac{1}{2}(p^2-m^2) \right)~.
	\label{scalarLagrangian}
\end{align}
The system is invariant under the following gauge transformations 
\begin{align}
	\delta x_{\mu}=\xi p^{\mu}~, \qquad \delta p_{\mu}=0~, \qquad \delta e=\dot{\xi}~,
	\label{scalargauge}
\end{align}
generated by the first-class constraint 
 \begin{align}
 	Q_0\equiv \frac{1}{2} \left(p^2-m^2\right)~.
 \end{align}
Following the Dirac procedure, 
and equipped with the Hamiltonian $H=e Q_0$,
the quantization consists in defining the Hilbert space as the linear space spanned by $|x\rangle$ or $|p\rangle$ where the physical states $|\psi \rangle$
satisfy 
$
	Q_0|\psi \rangle=0
$.
 
In this language, the free Feynman propagator $\langle \phi(x_f)\phi(x_i)\rangle$ of the corresponding scalar field $\phi(x)$
can be defined by the
matrix elements of the first-quantized operator
$(2Q_0+i\epsilon)^{-1}$, with 
the following path integral representation
\begin{align}
	\langle \phi(x_f)\phi(x_i)\rangle \equiv 
	\langle x_f|(2Q_0+i\epsilon)^{-1}|x_i\rangle
	&=\frac{1}{2}
	\int_{x(0)=x_i}^{x(1)=x_f} {\cal D}e {\cal D}x {\cal D}p\,
	e^{-i\int_0^1 dt\,\left(
	 p\cdot \dot{x}
	-e(Q_0+i\epsilon)
	\right)}~,
	\label{scalar}
\end{align}
where the integration measures have been normalized to unity. 
The validity of \cref{scalar} is perhaps clearer after gauge fixing 
$e(t)=T$. Thanks to $\delta e/\delta \xi$ in \cref{scalargauge} being field-independent, this choice yields a trivial Faddeev-Popov determinant. Then, 
the path integral over the einbein $e(t)$ reduces to an integration over 
gauge-non-equivalent parametrizations labeled by the Schwinger proper time $T$. After rescaling $t\to Tt$ in the action, 
the r.h.s. of \cref{scalar} becomes
\begin{align}
	\frac{1}{2}\int_0^{\infty} dT &\int_{x(0)=x_i}^{x(T)=x_f}
	{\cal D}x {\cal D}p\,
	e^{-i\int_0^T dt\,\left(
		p\cdot \dot{x}
		-Q_0-i\epsilon
		\right)}
	=		\frac{1}{2}\int_0^{\infty} dT
	\langle x_f|e^{i(Q_0+i\epsilon)T}|x_i\rangle
	~,
	\label{scalarT}
\end{align}
which matches $\langle x_f|(2Q_0+i\epsilon)^{-1}|x_i\rangle$ by virtue of \cref{schwinger}.
%
%

The procedure above can be easily generalized to a scalar particle propagating in a classical background $B(x)$. In that case one has to isolate the quadratic part in the field theory Lagrangian
 ${\cal L}^{(2)}=\phi^{*}(x)\Delta(x_{\mu},\partial_{\mu})\phi(x)$. Then,  
the corresponding first-quantized system exhibits again a gauge symmetry 
generated by the Noether charge 
\begin{align}
Q_0^B(\hat x, \hat p)\equiv\frac{1}{2}\Delta(\hat x_{\mu},-i\hat p_{\mu})	~.
\end{align}
Thus,  the definition in  \cref{scalar} is still valid, after replacing $Q_0\to Q_0^B$ in Weyl-ordered form. 

\begin{figure}
	\centering
	\includegraphics[width=0.5\linewidth]{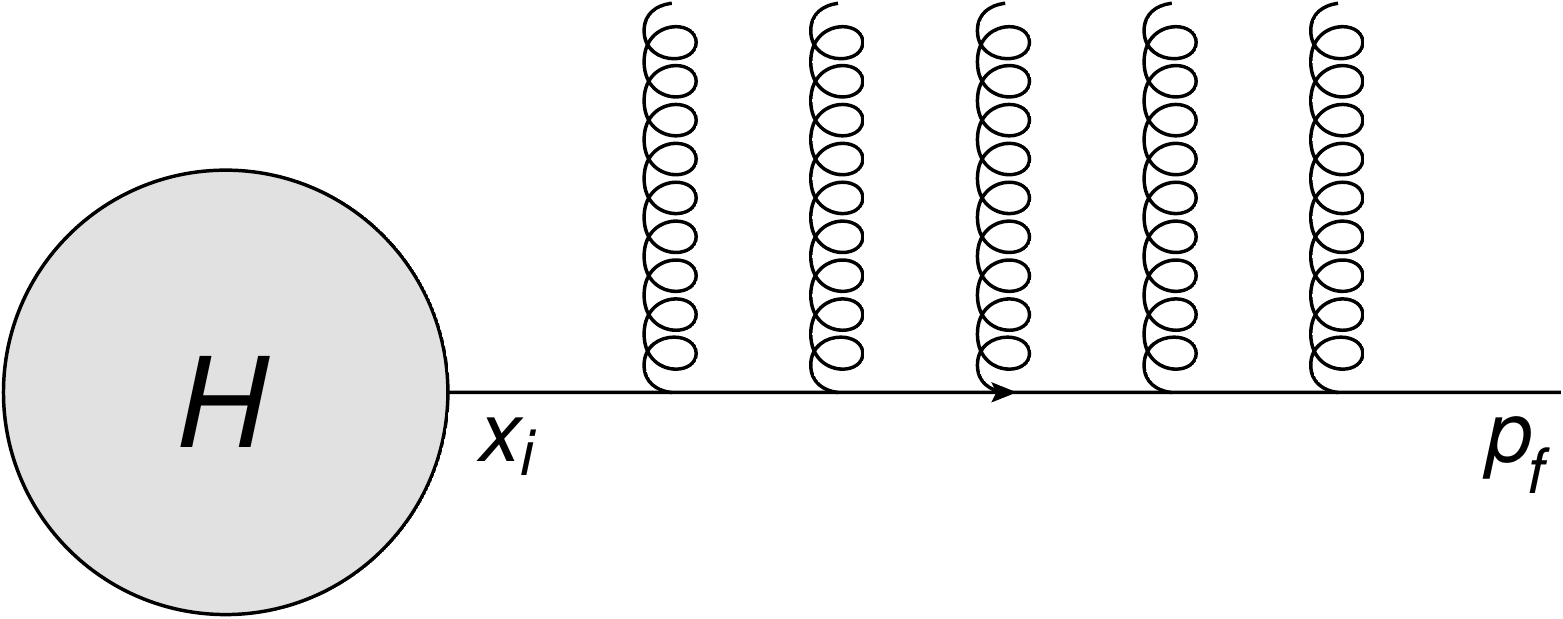}
	\caption{Schematic depiction of a propagator dressed by soft radiation, 
		 from 
		an initial state
	localized at position $x_i$ in the Hard function $H$ to a final state of momentum $p_f$.}
	\label{fig:xp}
\end{figure}

So far we have considered a dressed propagator between an initial and a final spacetime points $x_i$ and $x_f$, since this corresponds to the two-point function $\langle \phi(x_f)\phi(x_i)\rangle$ in the common second-quantized approach.  
On the other hand, in the first-quantized approach nothing prevents us from working 
in a mixed position-momentum representation. For instance, 
 we can consider
\begin{align}
	\langle p_f|(2Q_0^B+i\epsilon)^{-1}|x_i\rangle&= \frac{1}{2}\int_0^{\infty}dT\, \langle p_f|e^{-i(Q_0^B (\hat x, \hat p)+i\epsilon) T}|x_i\rangle \notag \\
	&=\frac{1}{2}\int_0^{\infty}dT\, \int_{x(0)=x_i}^{p(T)=p_f} {\cal D}x{\cal D}p\, e^{ip(T)\cdot x(T)-i\int_0^{T}dt\,(p\cdot \dot x -Q_0^B (  x,   p)-i\epsilon)} ~,
\label{mixed}
\end{align}
where the uncommon 
term $p(T)\cdot x(T)$ is due to this mixed boundary conditions. 
This is the representation used in \cite{Laenen:2008gt} for asymptotic propagators
 of a scattering amplitude. Indeed, as depicted in \cref{fig:xp}, in this case
one typically considers a 
dressed propagator emitted from the hard function at a spacetime point $x_i^{\mu}$, which will be integrated over, to a final state of momentum $p_f^{\mu}$.
The reason for having a final state with well-defined momentum 
 is that
eventually we would like to perform the path integral over $x^{\mu}$ order by order in the soft expansion w.r.t. the hard momentum $p_f^{\mu}$.

The previous arguments become clear after
expanding around the classical
solutions
\begin{align}
	p^{\mu}(t)=p_f^{\mu}+\tilde{p}^{\mu}(t)~, \qquad x^{\mu}(t)=x_i^{\mu}+p_f^{\mu}t+\tilde x^{\mu}(t)  ~.
	\label{eom}
\end{align}
Then, the dressed propagator reads
\begin{align}
	\frac{\langle p_f|(2Q_0^B+i\epsilon)^{-1}|x_i\rangle}{\langle p_f|x_i\rangle}
		&=\frac{1}{2}
	\int_0^{\infty}dT\, \int_{\tilde x(0)=0}^{\tilde p(T)=0} {\cal D}\tilde x{\cal D}\tilde p\, e^{
		-i\int_0^{T}dt\,\left(\tilde p\cdot p_f+\tilde p\cdot\dot{\tilde x} - Q_0^B (  x(\tilde x),p(\tilde p))-i\epsilon\right)} ~,
	\label{tildexp}
\end{align}
where we have included the normalization factor $\langle p_f|x_i\rangle=e^{ip_f\cdot x_i}$. The need for such a factor is evident by considering the free case that should return $i(p_f^2-m^2)^{-1}$.

Before continuing with the calculation of \cref{tildexp}, we discuss a point that will 
become central
for the spinning case in the next sections. Looking at \cref{tildexp}, one is tempted to pull $Q_0^B$
out of the integration over $t$, since
it is a conserved charge. Of course, this is true only on the equations of motion, therefore the correct statement 
is formulated  in terms of its expectation value, i.e.
\begin{align}
	\int_0^T dt   \langle Q_0^B(x(t),p(t))\rangle= \langle Q_0^B(x(t_0),p(t_0))\rangle T~,
	\label{conserved}
\end{align}
where $t_0$ is arbitrary. 
However, this is different from $\langle (Q_0^B)^{-1}\rangle$
which appears on the l.h.s. of \cref{tildexp}. Therefore,
the only thing we could do on the same line of reasoning is to note that
\begin{align}
\frac{d}{dt}\langle (Q_0^B)^{-1}\rangle
=-\langle \dot{Q}_0^B (Q_0^B)^{-2}\rangle~.
\label{inverse}
\end{align}
Then, a simple calculation from Noether's theorem reveals 
that a generic correlator of some operator $F[q_i]$ with
the time derivative of a Noether charge $\dot{Q}[q_i]$ can be expressed in terms of the 
  transformed canonical variables $\delta q_i(t)$ as\footnote{We are assuming that the symmetry 
 	is non-anomalous.}
\begin{align}
\langle \dot{Q}[q_i] F[q_i] \rangle
=-i \sum_i \langle \delta q_i(t) 
\frac{\delta F[q_i]}{\delta q_i(t)} \rangle
~.
\end{align}
Therefore, 
\begin{align}
	\frac{d}{dt}\langle (Q_0^B)^{-1}\rangle
	&=-i  \langle \delta x_{\mu}(t) 
	\frac{\delta (Q_0^B)^{-2}}{\delta x_{\mu}(t)} \rangle 
	-i  \langle \delta p_{\mu}(t) 
	\frac{\delta (Q_0^B)^{-2}}{\delta p_{\mu}(t)} \rangle 
	~.
	\label{ward}
\end{align}
In the free case, the Noether charge 
$Q_0=\frac{1}{2}(p^2-m^2)$ does not depend on $x$ and the momentum is gauge invariant, i.e. $\delta p=0$. Hence, the r.h.s. 
of \cref{ward} vanishes and one can effectively evaluate 
$Q_0(x(t),p(t))$ in the exponent of \cref{tildexp}
at an arbitrary time, say $T$. 	  Then, the remaining integrations over $\tilde p^{\mu}$ and 
$\tilde x^{\mu}$ are trivial, and one gets
the free Feynman propagator with momentum $p_f^{\mu}$.

However, for a generic dressed propagator, the charge 
$Q_0^B$ \emph{does}
depend on
$x^{\mu}$ and the momentum $p^{\mu}$ is \emph{not} gauge invariant. This makes \cref{ward} more involved, 
so that it is actually
more convenient to keep $Q_0^B$ inside the integral over $t$. 
To see how to proceed,  we consider in the next section a specific case
for the background field $B(x)$.

\subsection{Asymptotic propagators in a gauge boson background}
We consider a background gauge boson field
$A^{\mu}$ that for simplicity we assume to be abelian. In this
 case, we define $Q_0^A$ from the quadratic part of the scalar QED Lagrangian as
 \begin{align} 
 	Q_0^A(\hat x,\hat p)\equiv \frac{1}{2}\left((\hat p_{\mu}-A_{\mu}(\hat x))^2-m^2\right)
=\frac{1}{2} \hat p^2-\hat p\cdot A(\hat x)-i\frac{1}{2}(\partial\cdot A(\hat x))+\frac{1}{2}A^2(\hat x)~,
 	\label{scalarNoether}
\end{align}
where in the second equality we took into account that
the path integral representation in \cref{mixed} requires a 
Weyl-ordered Hamiltonian.

Then, we can plug \cref{scalarNoether} into \cref{tildexp} and perform the Gaussian integration over $\tilde p$.
We get
\begin{align}
	\frac{\langle p_f|(2Q_0^A+i\epsilon)^{-1}|x_i\rangle}{\langle p_f|x_i\rangle}
	&= \frac{1}{2}
	\int_0^{\infty}dT\,e^{i\frac{1}{2}(p_f^2-m^2+i\epsilon)T}
	f(x_i,p_f,T)   ~,
	\label{dressed}
\end{align}
where we have defined
\begin{align}
	f(x_i,p_f,T)&=
	\int_{\tilde x(0)=0} {\cal D} \tilde x\, 
	{\cal P} \exp\Bigg[ i\int_0^{T}dt\,\frac{1}{2}\dot{ \tilde x}^2 
	+(p_f+\dot{ \tilde x})\cdot A(x_i+p_f t + \tilde x(t)) 
	\notag \\
	&\qquad +\frac{i}{2}\partial\cdot A(x_i+p_f t+ x(t)) \Bigg] ~.
	\label{ftau}
\end{align}
We see that \cref{dressed} represents a dressed propagator in terms of a radiative factor $f$, 
equal to unity in the free case, 
which takes into account the interactions with the background field via 
four one-dimensional fields $x^{\mu}(t)$ living on the worldline of proper time $T$. 

Now we define the \emph{asymptotic} dressed propagator as the dressed propagator
truncated of the external free propagator of momentum $p_f$. This means that 
we should consider
\begin{align}
	&i(p_f^2-m^2) \frac{\langle p_f|(2Q_0^A+i\epsilon)^{-1}|x_i\rangle}{\langle p_f|x_i\rangle}
	=
	\int_0^{\infty}dT\,\left(\frac{d}{dT}e^{i\frac{1}{2}(p_f^2-m^2+i\epsilon)T}\right)
	f(x_i,p_f,T) ~.
	\label{asymptotic}
\end{align}
Assuming that the dressed propagator in \cref{dressed} develops a simple pole
for ${p_f^2\to m^2}$ with residue one and that the factor $f(x_i,p_f,T)$ remains finite in this
limit, we can integrate by parts to get
\begin{align}
	\lim_{p_f^2\to m^2}&i(p_f^2-m^2) \frac{\langle p_f|(2Q_0^A+i\epsilon)^{-1}|x_i\rangle}{\langle p_f|x_i\rangle}
	=	\lim_{p_f^2\to m^2}
	 f(x_i,p_f,\infty) ~.
	 \label{limits}	
\end{align}
Therefore, the asymptotic dressed propagator equals \cref{ftau} in the limit $T\to \infty$.
We are now ready to perform the remaining path integration in the soft expansion.

Following \cite{Laenen:2008gt}, we introduce a book-keeping parameter $\lambda$  and rescale 
$p_f^{\mu}\to \lambda n^{\mu}$, such that the soft expansion corresponds to an 
expansion in $1/\lambda$. 
Accordingly, it is convenient
to re-define the integration variable in \cref{ftau} as $t\to t/\lambda$.
Then, 
we get
\begin{align}
	f(x_i,n\lambda,\infty)=
	& 	\int_{\tilde x(0)=0} {\cal D} \tilde x\, 
	e^{i\int_0^{\infty}dt\,\left(\frac{\lambda}{2}\dot{ \tilde x}^2
		+(n+\dot  {\tilde x})\cdot A(x_i+n t + \tilde x(t))+\frac{i}{2\lambda}\partial \cdot A(x_i+n t+ \tilde x(t))\right)}  ~.
	\label{1qft}
\end{align}
The crucial observation is that all two-point correlators of $\tilde x^{\mu}$ and $\dot{\tilde x}^{\mu}$ are of order $1/\lambda$ and thus at a given order in $1/\lambda$
we need to include a finite number of diagrams. Therefore, by Taylor expanding $A_{\mu}$ in powers
of $\tilde x^{\mu}$ we can solve this one-dimensional QFT order by order in $1/\lambda$ and to all-orders in the coupling constant. 
More specifically, 
up to order $1/\lambda^m$ we need the diagrams with at most $m$ propagators, hence with vertices with at most $2m$ powers of $\tilde x^{\mu}$. Then, the sum of all diagrams can be rearranged in the exponential of the sum of connected diagrams.
The calculation has been carried out in detail in \cite{Laenen:2008gt}. For the sake of completeness, the most important steps are reviewed in 
\cref{softexpansion}. The upshot is that
the asymptotic dressed propagator $f(x_i,p_f,\infty)$ for a scalar particle reduces at NLP to the Generalized Wilson line defined in \cref{wilson} with $J^{\mu\nu}=L^{\mu\nu}$.

The derivation we have reviewed in this section can be generalized to the case of a non-abelian gauge field. In this case the dressed propagator becomes matrix-valued, hence we have two routes: either we introduce additional Grassmann variables on the worldline (as discussed in\cite{Ahmadiniaz:2015xoa}) or we stick with matrices in the exponent after introducing a path ordering prescription. Although the former approach is more elegant and the quantization of the model is derived from first principles, the latter
is often preferred in practical calculations involving soft gluons and it is the choice adopted in \cite{Laenen:2008gt}. Therefore, also in this work we stick with the second option. Then, the definition of \cref{wilson} is essentially the same, although the exponentiation is not derived in terms of connected diagrams but the so-called webs \cite{Laenen:2008gt, Gardi:2010rn, Gardi:2013ita, Agarwal:2020nyc}.

\section{Spin one-half}\label{dirac}

Having reviewed the generalized Wilson line in the scalar case, 
we are going to present the Dirac case following the same procedure i.e. starting
from the classical single-particle model and its symmetries. 
This will bring us to identify the distinctive features of asymptotic propagators and
subsequently to justify \cref{wilson} for spin $1/2$.

\subsection{Worldline representation}
We start with the phase space action for a free massless spin $1/2$ particle, which reads
\begin{align}
	S=~\int dt\,\left( p_{\mu}\dot{x}^{\mu}
	+  \frac{i}{2}\psi^{\mu}\dot{ \psi_{\mu}}
	-H\right)~,
	\label{freeDirac}
\end{align}
where the Hamiltonian $H$ is
\begin{align}
	H&=\frac{1}{2}ep_{\mu}p^{\mu}
	+i\chi\psi_{\mu}p^{\mu}~.
	\label{hamiltonian}
\end{align}
Here $\psi_{\mu}$ are classical Grassmann variables,
 which after quantization
satisfy the Clifford algebra $\{\hat\psi_{\mu},\hat\psi_{\nu}\}=\eta_{\mu\nu}$.
Unlike the scalar case, this action enjoys two gauge symmetries: local supersymmetry and reparametrization invariance, with the respective gauge bosons $e(t)$ and $\chi(t)$.
The transformations are
\begin{align}
	\delta x^{\mu}=\xi p^{\mu}+i\zeta \psi^{\mu}~, \qquad \delta p^{\mu}=0~,
	 \qquad \delta \psi^{\mu}=-\zeta p^{\mu}~,
	 \qquad \delta e=\dot{\xi}~,  \qquad \delta \chi=\dot{\zeta}~,
\end{align}
while
the relative Noether charges are
\begin{align}
	Q_0\equiv \frac{1}{2}p^2~, \qquad Q_1\equiv \psi\cdot p~,
	\label{charges}
\end{align}
whose kernels define the physical Hilbert space. 
More specifically, the Dirac equation $Q_1|\psi\rangle=0$
generates the other constraint $Q_0|\psi\rangle=0$
thanks to the $N=1$ supersymmetry algebra $\{Q_1,Q_1\}=-2iQ_0$.

Consequently, the Feynman propagator for the field $\psi(x)$ admits the following 
first-quantized representation
\begin{align}
&\langle p_f|\frac{Q_1}{2Q_0+i\epsilon}|x_i \rangle
\notag \\&=
\int {\cal D} e  {\cal D} \chi
\int_{\psi(1)=\Gamma} {\cal D}\psi
\int_{x(0)=x_i}^{p(1)=p_f}  {\cal D}x {\cal D}p \,
e^{ip(1)\cdot x(1)-i\int_0^1 dt\,\left(
	p\cdot \dot{x}+\frac{i}{2} \psi \cdot \dot{ \psi}
	-eQ_0-\chi Q_1-i\epsilon
	\right)}~.
	\label{freeDirac2}
\end{align}
A few comments about the boundary conditions are in order. First,
we note that as for the scalar case we are considering
a mixed representation from an initial 
state of position $x_i^{\mu}$ (which eventually is integrated over) to a final state
with momentum $p_f^{\mu}$. Then, we note that
 the propagator is open and thus we do not set antiperiodic boundary conditions
$\psi_{\mu}^{\mu}(0)+\psi_{\mu}^{\mu}(1)=0$, as typically done for the computation of the effective action. In fact, it is known (see e.g. \cite{Alexandrou:1998ia, Ahmadiniaz:2020wlm}) that an open propagator requires 
the inhomogeneous conditions $\psi_{\mu}^{\mu}(0)+\psi_{\mu}^{\mu}(1)=\Gamma^{\mu}$,
where $\Gamma^{\mu}$ is a set of constant Grassmann variables that should generate the spin structure of the propagator. Given that the external states of a scattering amplitude must have a well-defined spin, while we sum over the spin values of the initial state attached to 
the hard function, we set only the final value of the spin variable by requiring $\psi_{\mu}(1)=\Gamma_{\mu}$, where $\Gamma_{\mu}$ 
  will eventually
be set proportional to the gamma matrices.
 
Then we can proceed as in the scalar case and gauge-fix the Lagrange multipliers, by setting $(e(t),\chi(t))=(T,\theta)$. The corresponding path integrations become regular integrals over the proper time $T$ and the ``supertime'' $\theta$, respectively. In this way, \cref{freeDirac2} reads
\begin{align}
&\langle p_f|\frac{Q_1}{2Q_0+i\epsilon}|x_i \rangle
\notag \\&= \frac{1}{2}
	\int_0^{\infty} dT \int d\theta
	\int_{\psi(T)=\Gamma} {\cal D}\psi
	\int_{x(0)=x_i}^{p(T)=p_f}  {\cal D}x {\cal D}p \,
	e^{ip(T)\cdot x(T)-i\int_0^T dt\,\left(
		p\cdot \dot{x}+\frac{i}{2} \psi \cdot \dot{ \psi}
		-Q_0-\frac{\theta}{T}Q_1-i\epsilon
		\right)}~.
	\label{fixedDirac}
\end{align}
Once again, the role of the Lagrange multipliers is to exponentiate the constraints $Q_0$ and $Q_1$. 
In particular, the Grassmann nature of the supertime $\theta$ allows the exponentiation of the numerator of the propagator, while the proper time $T$ exponentiates the denominator as for the scalar case.

The generalization to the presence of an abelian gauge boson background field is straightforward. This can be achieved by replacing $p_{\mu}\to p_{\mu}-A_{\mu}\equiv \Pi_{\mu}$
in the charge $Q_1$, while the charge $Q_0$ is obtained by the supersymmetry algebra
$\{Q_1,Q_1\}=-2iQ_0$. The new transformations read
\begin{align}
&	\delta x^{\mu}=\xi \Pi{\mu}+i\zeta \psi^{\mu}~, \qquad\qquad \delta p^{\mu}=
	-\frac{\xi}{2}\frac{\delta \Pi^2}{\delta x_{\mu}}-i\zeta \psi_{\nu}\frac{\delta \Pi^{\nu}}{\delta x_{\mu}}~,
	\qquad\qquad \delta \psi^{\mu}=-\zeta \Pi^{\mu}~, \notag \\
&	\qquad \qquad\qquad\qquad\delta e=\dot{\xi}+2i\chi \zeta~,\qquad\qquad  \qquad \delta \chi=\dot{\zeta}~,
\end{align}
and are generated by the new charges
\begin{align}
	Q_0^A\equiv \Pi^2+\psi_{\mu}\psi_{\nu}F^{\mu\nu}~, \qquad Q_1^A\equiv\psi\cdot \Pi~.
\end{align}
Then, the structure of \cref{fixedDirac} remains the same.

At this point, we note that unlike the scalar case where the dressed propagator 
is represented by the expectation values of the \emph{inverse} of a conserved charge, 
here we have an additional charge in the numerator.
Therefore, we can directly apply \cref{conserved} 
and evaluate $Q_1^A$ in the exponent of \cref{fixedDirac} at an arbitrary time
and pull it out of the integral. 
Given the boundary conditions that fix $\psi(T)$, we choose this arbitrary time to be $T$.
Therefore we consider
\begin{align}
	\int_0^T dt   \langle Q_1^A(x(t),p(t))\rangle= \langle Q_1^A(x(T),p_f)\rangle T~,
	\label{conserved2}
\end{align}
and \cref{fixedDirac} becomes
 \begin{align}
 	&\langle p_f|\frac{Q_1^A}{2Q_0^A+i\epsilon}|x_i \rangle
 	\notag \\&=\frac{1}{2}
 	\int_0^{\infty} dT \int d\theta
 	\int_{\psi(T)=\Gamma} {\cal D}\psi
 	\int_{x(0)=x_i}^{p(T)=p_f}  {\cal D}x {\cal D}p \,
 	e^{ip(T)\cdot x(T)+i \theta Q_1^A(x(T),p_f)-i\int_0^T dt\,\left(
 		p\cdot \dot{x}+\frac{i}{2} \psi \cdot \dot{ \psi}
 		-Q_0^A-i\epsilon
 		\right)}~.
 	\label{conservedDirac}
 \end{align}
This is the worldline representation for a Dirac propagator in the presence of an abelian background field with boundary conditions suitably chosen to handle the asymptotic states of a scattering amplitudes. 
In the following sections we are going to discuss how such representation is related to the numerator and the denominator of a dressed propagator, respectively. 

\subsection{Numerator contribution}
\label{sec:num}
The presence of $x(T)$ in the argument of the numerator contribution
$Q_1^A$ in \cref{conservedDirac}
 seems to suggest that the procedure outlined above is pointless, since
 $x(T)$ is not fixed by the boundary conditions.
 However, there are at least two cases where this can be handled: the free 
 case
 and, most importantly, the asymptotic case. We start with the former.
 
 In the free case the gauge field vanishes so the Noether charges $Q_0$ and $Q_1$ do not depend
 on $x_{\mu}$. Making use of the boundary conditions on $p_{\mu}$ and $\psi_{\mu}$ we get
 \begin{align}
 	Q_1(x(T),p(T))=\Gamma\cdot p_f~.
 \end{align}
This gives
\begin{align}
	&\langle p_f|\frac{Q_1}{2Q_0+i\epsilon}|x_i \rangle
	\notag \\&= e^{ip_f\cdot x_i}\frac{1}{2}
	\int_0^{\infty} dT \, e^{\frac{i}{2}(p_f^2+i\epsilon)T}\int d\theta \,e^{i\theta\, \Gamma\cdot p_f}
	\int_{\psi(T)=\Gamma} {\cal D}\psi\,e^{\int_0^{T} dt\,\frac{1}{2} \psi \cdot \dot{ \psi}}
	\int_{\tilde{x}(0)=0}^{\tilde{p}(T)=0}  {\cal D}\tilde{x} {\cal D}\tilde{p} \,
	e^{i\int_0^T dt\,\left(
		\frac{\tilde{p}^2}{2}+\tilde{p}\cdot \dot{\tilde{x}}
		\right)}~.
	\label{free}
\end{align}
 The remaining path integrals are Gaussian and can be 
 absorbed together with $e^{ip_f\cdot x_i}$ into the normalization factor  $\langle p_f|x_i \rangle$. Then, assuming that the constant $\Gamma_{\mu}$ can be represented
 by the Dirac gamma matrices $\gamma_{\mu}$, one is left with
 \begin{align}
 \langle p_f|x_i \rangle^{-1}	\langle p_f|\frac{Q_1}{2Q_0+i\epsilon}|x_i \rangle=
 \frac{1}{2}
 	\int_0^{\infty} dT\, e^{\frac{i}{2}(p_f^2+i\epsilon)T}\int d\theta\,
 	 e^{i \slashed p_f\theta}
 	 =\frac{i \slashed p_f}{p_f^2+i\epsilon}~,
 	\label{fixedDirac4}
 \end{align}
 as expected. 

Coming back to the interacting case in \cref{conservedDirac}, we first note
 that the Noether charge 
 with the given boundary conditions reads
 \begin{align}
	Q_1^A(x(T),p(T))= \Gamma \cdot (p_f-A(x(T)))~.
	\label{chargeAsympt}
\end{align}
Then, it is convenient to get rid of the Gaussian integration over $p_{\mu}$, to get
\begin{align}
 \langle p_f|x_i \rangle^{-1}	\langle p_f|\frac{Q_1^A}{2Q_0^A+i\epsilon}|x_i \rangle=
\frac{1}{2}
\int_0^{\infty} dT\, e^{\frac{i}{2}(p_f^2+i\epsilon)T}\int d\theta\,
e^{i \Gamma \cdot p_f\theta}
	f(x_i,p_f,\Gamma,T,\theta)   ~,
	\label{dressedDir}
\end{align}
where, in analogy with \cref{ftau}, we have defined
\begin{align}
	f(x_i,p_f,\Gamma,T,\theta)&=
		\int_{\psi(T)=\Gamma} {\cal D}\psi\, 
	\int_{\tilde x(0)=0} {\cal D} \tilde x\, 
	e^{ -i\theta \psi\cdot A(x_i+p_f T+ \tilde{x}(T))}\notag \\
	&\qquad e^{i\int_0^{T}dt\,\left(-\frac{i}{2}\psi \cdot \dot{ \psi}+\frac{1}{2}\dot{ \tilde x}^2 
	+(p_f+\dot{ \tilde x})\cdot A(x_i+p_f t + \tilde x(t)) 
 +\frac{i}{2}\partial\cdot A(x_i+p_f t+ \tilde{x}(t))+\frac{1}{2}\psi_{\mu}\psi_{\nu}F^{\mu\nu}\right)} ~.
	\label{ftauDirac}
\end{align}

Now, for asymptotic propagators we should truncate the external line
by multiplying by a free inverse propagator, similarly to \cref{asymptotic}.
Hence, we should consider
\begin{align}
	\bar u(p_f)\frac{i}{\slashed{p_f}}p_f^2
	\langle p_f|x_i \rangle^{-1} &	\langle p_f|\frac{Q_1^A}{2Q_0^A+i\epsilon}|x_i \rangle
	\notag \\
	&=	\bar u(p_f)
	\int_0^{\infty} dT\, \frac{d}{dT}\left(e^{\frac{i}{2}(p_f^2+i\epsilon)T}\right)
	\int d\theta\,
	e^{i \Gamma \cdot p_f\theta}
	f(x_i,p_f,\Gamma,T,\theta)  ~.
	\label{asymptoticDirac}	
\end{align}
With similar assumptions leading to \cref{limits}, we can integrate \cref{asymptoticDirac} by parts to get 
\begin{align}
\bar u(p_f)\frac{i}{\slashed{p_f}}p_f^2
\langle p_f|x_i \rangle^{-1}	\langle p_f|\frac{Q_1^A}{2Q_0^A+i\epsilon}|x_i \rangle
	= 	\lim_{p_f^2\to 0} \bar u(p_f) \frac{1}{\slashed{p_f}}
	\int d\theta\,
	e^{i \Gamma \cdot p_f\theta}
	f(x_i,p_f,\Gamma,\infty,\theta) ~.
	\label{limitsDirac}	
\end{align}
Therefore, also in the Dirac case, the asymptotic propagator is obtained by taking 
the $T\to\infty$ limit of the dressed propagator, in this case of \cref{ftauDirac}.

The practical consequence  is that the gauge field $A_{\mu}(x(T))$ in \cref{chargeAsympt}
is evaluated at infinity. 
Let us assume for a moment that it vanishes.
 Then, we can drop the $\theta$-dependence
  in $f(x_i,p_f,\Gamma,\infty,\theta)$ and the $\theta$-integral in  
\cref{limitsDirac} cancels with $\slashed{p}_f$ in the denominator. 
Therefore, the asymptotic dressed propagator in the presence
of an asymptotically vanishing background field reads
\begin{align}
	\lim_{p_f^2\to 0} \bar u(p_f)
	f(x_i,p_f,\Gamma,\infty) ~.
	\label{limitsDirac2}	
\end{align}
In other words, the numerator $Q_1^A$ of the dressed propagator has been taken effectively free,
leaving the entire dependence on the background field in the denominator $Q_0^A$. This is very reminiscent 
of the closed loop topology of the one-loop effective action, as pointed out in the introduction.

At this point we should examine more carefully the assumption of the vanishing $A_{\mu}(x(T))$.
In fact, what we should consider is the asymptotic limit in the soft expansion.
 We have seen in the scalar case that such expansion
is achieved by rescaling $p_f^{\mu}\to \lambda n^{\mu}$ and then expanding in $1/\lambda$.
Looking at \cref{chargeAsympt} we can immediately see that $A_{\mu}(x(T))$ is subleading w.r.t. $p_f^{\mu}$.
This is reassuring, since it is well-known that one should get a regular Wilson line
in the strict soft limit, which is insensitive to the spin of the emitter. The question is whether $A_{\mu}(x(T))$ can also be neglected
at subleading power in the soft expansion. 

The answer is yes. This can be seen by
mimicking what done in \cref{softexpansion} for the scalar case: after 
expanding the gauge
field in powers of $\tilde x^{\mu}$, the only relevant
vertex up to order $1/\lambda$ is 
 \begin{align}
	\Gamma \cdot A(p_f T)
	=\Gamma^{\mu} 
	\int \frac{d^dk}{(2\pi)^d} e^{ik\cdot p_f T} \tilde{A}_{\mu}(k) 
	~.
	\label{oscillates}
\end{align}
This contains no power of $\tilde x^{\mu}$ and can thus be pulled out
of the path integral. Vertices with higher powers of $\tilde x^{\mu}$ would be needed in diagrams
with $\tilde x$-propagators, which are thus subleading in $1/\lambda$.

The integral in \cref{oscillates} is suppressed if
the integrand oscillates very rapidly, as happens
in the asymptotic limit $T\to\infty$, assuming  an integrable 
$\tilde{A}_{\mu}(k)$. 
An obvious counter-example is given by the limiting case of a constant
 field, where
 the integrand has support only at $k=0$.  
 One might wonder
 whether the soft limit is dangerous here since it corresponds to
 a long-wavelength background field with 
 $\tilde{A}_{\mu}(k)$ concentrated around $k\to 0$. However, 
 the soft limit on the worldline has been defined
 by rescaling $p_f\to \lambda n$ and letting $\lambda\to\infty$
 so that $p_f\cdot k$ does not tend to zero. In other words, while
 it is legitimate to assume a constant field strength tensor $F^{\mu\nu}$ in this limit, the 
 field $A^{\mu}$ is not exactly constant and it vanishes for $T\to\infty$. 
  Therefore, \cref{oscillates} vanishes and we conclude 
  that 
  for an asymptotic propagator dressed by (next-to-)soft radiation
 the numerator contribution $Q_1^A$ reduces to the free numerator $Q_1$.

\subsection{Denominator contribution}
\label{sec:den}
Now that we have established that \cref{limitsDirac2} holds for asymptotic  propagators
dressed of soft radiation, 
we can perform the remaining path integrations in \cref{ftauDirac} in the limit $T\to\infty$.
In principle we could integrate out $\psi$ exactly, since it appears quadratically. 
However, this will bring an intricate dependence on the background 
field\footnote{The result
 significantly simplifies for a constant background field  \cite{Reuter:1996zm}.}. 
For our purposes it is actually more convenient to perform the integration
order by order in the soft expansion, where we can use  
the same rescaling implemented in the $x$-integration, i.e. $p_f\to\lambda n$ and $t\to t/\lambda$, to get
\begin{align}
	\int_{\psi(\infty)=\Gamma} {\cal D}\psi\, 
	& e^{i\int_0^{\infty}dt\,\left(-\frac{i}{2}\psi \cdot \dot{ \psi}
		+\frac{1}{2\lambda}\psi_{\mu}\psi_{\nu}F^{\mu\nu}\right)} ~.
	\label{ftauDirac2}
\end{align}

It is convenient to expand the field $\psi_{\mu}(t)$ around the boundary condition with 
the replacement $\psi(t)=\psi(T)+\widetilde\psi(t)$. In this way \cref{ftauDirac2} becomes
\begin{align}
	e^{\frac{i}{2\lambda}\Gamma_{\mu}\Gamma_{\nu}\int_0^{\infty}dt\,F^{\mu\nu}(x_i+nt+\tilde x(t))}
	\int_{\widetilde\psi(\infty)=0} {\cal D}\widetilde\psi\, 
	& e^{i\int_0^{\infty}dt\,\left(-\frac{i}{2}\widetilde\psi \cdot \dot{\widetilde \psi}
		+\frac{1}{2\lambda}\widetilde\psi_{\mu}\widetilde\psi_{\nu}F^{\mu\nu}(x_i+nt+\tilde x(t))\right)} ~.
	\label{ftauDirac3}
\end{align}
We note that the propagator $\langle\widetilde\psi(t)\widetilde\psi(t')\rangle$ is of order $\lambda^0$ and is proportional to the step function
 $\theta(t-t')$. The interaction term is of order $\lambda^{-1}$ and generates vertices 
 with various powers of $\tilde x$ by expanding $A_{\mu}(x(t))=A_{\mu}(x_i+n t+\tilde x(t))$
  around $\tilde x=0$. As already observed in the scalar case, the effect from $x_i\neq 0$
  is subleading in $1/\lambda$. Moreover, here this is multiplied by an additional 
  $1/\lambda$ coming from the vertex, hence it is a $1/\lambda^2$ effect that we can 
  neglect.  
Now we recall that the propagator of the $\tilde x(t)$ field 
(and more generally all correlators with
two powers of $\tilde x$ or $\dot{\tilde x}$) are of order $1/\lambda$. Therefore, vertices
with higher powers of $\tilde x$ are needed only for diagrams that are subleading
in the soft expansion. 
Thus we can expand
the gauge field in $F_{\mu\nu}$ at leading order in $\tilde x$
so that no dependence over $\tilde x$ is left and the path integral
over $\widetilde\psi$ in \cref{ftauDirac} decouples from the one over $\tilde x$.

In analogy with the free case of \cref{fixedDirac4}, we can absorb such remaining 
Gaussian integration over $\widetilde\psi$ in the normalization factor $\langle p_f|x_i \rangle$,
so that one is left with the factor  
\begin{align}
	\exp{\left(\frac{i}{2\lambda}\Gamma_{\mu}\Gamma_{\nu}\int_0^{\infty}dt\,F^{\mu\nu}(nt)\right)}
		&=	\exp{\left(\frac{i}{\lambda}S^{\mu\nu}
			\int \frac{d^dk}{(2\pi)^d} \,\frac{k_{\nu}}{n\cdot k}\, \tilde{A}_{\mu}(k)
			\right)} ~,
	\label{diagram}
\end{align}
where we used the representation $\Gamma_{\mu}=\frac{1}{\sqrt{2}}\gamma_{\mu}$
and we introduced the spin $1/2$ generator
 $S^{\mu\nu}=\frac{i}{4}[\gamma^{\mu},\gamma^{\nu}]$. Also, 
we set $x_i=0$ since it is a subleading effect, as discussed in \cref{softexpansion}. 
As we can see, \cref{diagram} equals the fourth term in the first line of \cref{wilson}.

The remaining path integral over $\tilde x$ matches exactly the scalar case and therefore we conclude
that the only difference between the Dirac and the scalar case is given by 
the result in \cref{diagram}, in agreement with \cite{Laenen:2008gt}. As we have remarked in the 
introduction with 
\cref{dressedDirac}, it is a pure ``denominator'' effect, which is present also in the effective action \cite{Strassler:1992zr}. It represents a chromo-magnetic interaction
between the emitter and the next-to-soft radiated particles by the 
coupling of $F^{\mu\nu}$ with the Lorentz generator $S^{\mu\nu}$.
This is precisely the spin term in the generalized Wilson line of \cref{wilson}.

So far we assumed an abelian background field. 
A similar derivation can be presented in the non-abelian case,
 by assuming that the exponentials are path-ordered. One still ends up with \cref{ftauDirac2}. However, in this case  
 $F^{\mu\nu}$  contains the commutator $[A^{\mu},A^{\nu}]$ which gives rise to the
 following spin-dependent term
  \begin{align}
	\frac{ig^2}{2\lambda}\int_0^{\infty} dt\, \Gamma^{\mu}\Gamma^{\nu}[A_{\mu}(nt),A_{\nu}(nt)]
	=\frac{ig^2}{\lambda} S^{\mu\nu}
	\int \frac{d^dk}{(2\pi)^d}\int \frac{d^dl}{(2\pi)^d} \,
	\frac{1}{n\cdot (k+l)} \tilde{A}_{\mu}(k)\tilde{A}_{\nu}(l) ~,
	\label{diagram2}
\end{align}
 where we exploited the anti-symmetry 
 of the Lorentz generator.
As we can see, \cref{diagram2} reproduces the spin-dependent term  
 with two gauge fields in \cref{wilson}.

We conclude this section with a small remark about the introduction
of a mass term, which is notoriously 
 subtle for Dirac particles. 
Traditionally, this requires an additional spin variable $\psi_{5}$ \cite{Brink:1976uf, Berezin:1976eg, Fradkin:1991ci, vanHolten:1995ds, Ahmadiniaz:2020wlm}.
This is a necessary choice if one wants to incorporate the mass term in the extended Hamiltonian of \cref{hamiltonian}, since the additional variable would make the term Grassmann-even.
Alternatively, it is sufficient to leave it out of the path integral as a Grassmann-odd projector, as already proposed in \cite{DiVecchia:1979gx, Alexandrou:1998ia}.
However, for the purpose of this paper
one is forced to include this extra dimensional variable 
in order to justify the vanishing contribution of the gauge field in the numerator
for asymptotic propagators. This can be achieved with 
 the conserved charges
\begin{align}
	Q_0^A\equiv \Pi^2+\psi_{\mu}\psi_{\nu}F^{\mu\nu}-m^2~, 
	\qquad Q_1^A\equiv\psi\cdot \Pi+m\psi_5~.
\end{align}
The derivation then is analogous: both the mass term in the numerator
and in the denominator can be factored out and cancel with the free inverse propagator.

\section{Spin one}
\label{sec:gluons}


\subsection{Generalized Wilson Line for gluons}
Before discussing the details
of the supersymmetric worldline model for higher spin particles, it is instructive 
to derive the GWL for gluons starting
from the dressed propagator of the corresponding field-theory, rather than 
the quantization of the relativistic particle.  
In fact, as we have seen in the Dirac case, the
  model with worldline fermions is introduced in this context
  to explain the role of the background field
 in the numerator of dressed propagators. 
 However, for spin-one we can use gauge invariance as a shortcut.
Indeed, recalling that the GWLs are meant to represent the external states of
a scattering amplitude,
 the dressed propagator can be 
computed in a gauge where the numerator is unity and thus does not 
depend on the background gauge field. 
Let us discuss this in more detail.

We start from the quadratic part of the field-theory Lagrangian.  
In order to preserve the gauge invariance w.r.t. the (soft) background 
field, it is convenient to work in the class of background-field-gauges.
This is a well-known procedure (see e.g. \cite{Schwartz:2013pla}), which consists of
replacing $A^{\mu}\to \widetilde A^{\mu}+A^{\mu}$ in the Yang-Mills Lagrangian
${\cal L}= -\frac{1}{4}\left(F_{\mu\nu}^a\right)^2$, 
and subsequently fix the gauge by adding the term 
\begin{align}
	{\cal L_{\text{gf}}}=-\frac{1}{2\xi}(\widetilde{D}_{\mu}A^{\mu})^2~,
\end{align}
with the corresponding ghost term
\begin{align}
	{\cal L_{\text{ghost}}}=(\widetilde{D}_{\mu}\bar\omega)^a(\widetilde{D}_{\mu}\omega)^a
	+gf^{abc}(\widetilde{D}_{\mu}\bar\omega)^aA_{\mu}^b \omega^c~.
\end{align}
Here, we defined
\begin{align}
	\widetilde D_{\mu}^{ab}=\partial_{\mu}\delta^{ab}-gf^{abc}\widetilde A_{\mu}^c~,
\end{align}
where we chose $\widetilde A^{\mu}$ to be the background field.
Importantly, this gauge-fixing term breaks the invariance only w.r.t. the propagating 
$A^{\mu}$ field, while the gauge symmetry of $\widetilde A^{\mu}$ is preserved.
Then, after some algebra, the quadratic part reads
\begin{align}
	{\cal L}_{A^2}=
	\frac{1}{2}A_{\mu}^a \left[
	\eta^{\mu\nu}
	(\widetilde{D}^{ab})^2
	-\left( 1-\frac{1}{\xi}\right)(\widetilde{D}^{\mu}\widetilde{D}^{\nu})^{ab}
	+igf^{abc}\widetilde F_{\rho\sigma}^c(S^{\mu\nu})^{\rho\sigma}
	\right]
	A_{\nu}^b~,
	\label{quadratic}
\end{align}
where we introduced the spin-one Lorentz generator 
\begin{align}
	(S_{\mu\nu})^{\rho\sigma}=i(\delta_{\mu}^{\rho}\delta_{\nu}^{\sigma}-\delta_{\nu}^{\rho}\delta_{\mu}^{\sigma})~.
	\label{gen}
\end{align}

Therefore, setting $\xi=1$ and taking the inverse of the expression in brackets in \cref{quadratic},
we see that the numerator of the dressed propagator is unity.
The denominator on the other hand has the same structure that we found in the spinor case,
i.e. a scalar term $D^2$ and a spin-dependent term $F_{\mu\nu}S^{\mu\nu}$ representing a chromo-magnetic interaction between the magnetic moment of the emitting particle and the background field. 
This leads us to define 
the following Hamiltonian
\begin{align}
	H^{\mu\nu,ab}&= \frac{1}{2}\left(\eta^{\mu\nu}
	({D}^{ab})^2
	+igf^{abc} F_{\rho\sigma}^c(S^{\mu\nu})^{\rho\sigma} \right)~,
\end{align}
where we have dropped the tilde over the background field.
Thus, by replacing $\partial_{\mu}\to-i\hat{p}_{\mu}$
the covariant derivative can be written as 
an operator in the Hilbert space generated by $\hat{x}$ and $\hat{p}$, 
i.e. 
\begin{align}
	D_{\mu}^{ab}(\hat{x},\hat{p})=-i\hat{p}_{\mu}\delta^{ab}+igA^{ab}_{\mu}(\hat x)~,	
\end{align}
where we defined as usual
\begin{align}
	A^{ab}_{\mu}=A^c_{\mu}T^{ab}_c=-if^{abc}A^c_{\mu}~, \qquad F^{ab}_{\mu\nu}=F^c_{\mu\nu}T^{ab}_c=-if^{abc}F^c_{\mu\nu}~.
\end{align}
Then, we can proceed as in the scalar case and consider
the following path integral representation for the dressed propagator
with a mixed position-momentum boundary conditions:
\begin{align}
	\langle p_f|(H_{\mu\nu}+i\epsilon)^{-1}|x_i\rangle&= \frac{1}{2}\int_0^{\infty}dT\, \langle p_f|e^{-i(H_{\mu\nu}(\hat x, \hat p)+i\epsilon) T}|x_i\rangle \notag \\
	&=\frac{1}{2}\int_0^{\infty}dT\, \int_{x(0)=x_i}^{p(T)=p_f} {\cal D}x{\cal D}p\, 
		{\cal P} e^{ip(T)\cdot x(T)\,\eta_{\mu\nu}-i\int_0^{T}dt\,(p\cdot \dot x\,\eta_{\mu\nu}
			 -H_{\mu\nu}(\hat x, \hat p)-i\epsilon)} ~,
	\label{mixedGauge}
\end{align}
where for notation purposes color indices have not been explicitly shown.
Performing the Gaussian integration over the momentum and factorizing the 
normalization factor $\langle p_f|x_i\rangle$, we get
\begin{align}
\frac{\langle p_f|(H_{\mu\nu}+i\epsilon)^{-1}|x_i\rangle}{\langle p_f|x_i\rangle}
	&= \frac{1}{2}
	\int_0^{\infty}dT\,e^{i\frac{1}{2}(p_f^2+i\epsilon)T}
	f_{\mu\nu}(x_i,p_f,T)   ~.
	\label{dressedGauge}
\end{align}
This expression can be compared with the corresponding representations in
\cref{dressed} and \cref{dressedDir}.
Here,  in analogy with \cref{ftau} and \cref{ftauDirac}, we defined
\begin{align}
	f_{\mu\nu}(x_i,p_f,T)&=
	\int_{\tilde x(0)=0} {\cal D} \tilde x\, 
	{\cal P} \exp\Bigg[ i\int_0^{T}dt\, \Big(
	\frac{1}{2}\dot{ \tilde x}^2 
	+(p_f+\dot{ \tilde x})\cdot A(x_i+p_f t + \tilde x(t)) 
	\notag \\
	&\qquad +\frac{i}{2}\partial\cdot A(x_i+p_f t+ \tilde x(t))\Big)\eta_{\mu\nu}
	+g(S_{\mu\nu})^{\rho\sigma}F_{\rho\sigma}(x_i+p_f t+ \tilde x(t))
	 \Bigg] ~,
	 \label{ftauGluon}
\end{align}
where once again we have expanded around the classical solutions given by \cref{eom}.

Since the numerator of the dressed propagator is proportional to $\eta_{\mu\nu}$, the 
truncation of the external propagator is harmless. Indeed, we get 
\begin{align}
	\epsilon^{*}_{\mu}(p_f)
	\,(i \,\eta^{\mu\rho}\,p_f^2 )\frac{\langle p_f|(H_{\rho\nu}+i\epsilon)^{-1}|x_i\rangle}{\langle p_f|x_i\rangle}
	&=\epsilon^{*\mu}(p_f)
	\int_0^{\infty}dT\,\left(\frac{d}{dT}e^{i\frac{1}{2}(p_f^2+i\epsilon)T}\right)
	f_{\mu\nu}(x_i,p_f,T) 
	~.
	\label{asymptoticGauge}
\end{align}
Integrating by parts and taking the on-shell limit $p_f^2\to 0$,
\cref{asymptoticGauge} reduces to 
\begin{align}
	\lim_{p_f^2\to 0} \epsilon^{*\mu}(p_f)\,\, f_{\mu\nu}(x_i,p_f,\infty)~.
	\end{align}
Therefore, in analogy with the scalar and the Dirac cases, the dressed
asymptotic propagator 
for a gluon
is given by \cref{ftauGluon} in the limit $T\to\infty$. 

Finally, the path integral can be solved order by order in the soft expansion
after replacing $p_f^{\mu}\to \lambda n^{\mu}$ and considering only diagrams up to order
$1/\lambda$, 
 as discussed in 
\cref{softexpansion} for the scalar case. 
Here, 
the only difference is given by the presence
of the $F_{\mu\nu}(S_{\mu\nu})^{\rho\sigma}$ term. 
By expanding the gauge field in $F_{\mu\nu}$
at leading order in $\tilde x^{\mu}$ we get two additional terms of order $1/
\lambda$. The first involves one
power of $A^{\mu}$:
  \begin{align}
 	\frac{ig}{\lambda}\int_0^{\infty} dt\, (S_{\mu\nu})^{\rho\sigma}\partial_{\rho}A_{\sigma}(nt)
 	=\frac{ig}{\lambda} (S_{\mu\nu})^{\rho\sigma}
 	\int \frac{d^dk}{(2\pi)^d} \,\frac{k_{\sigma}}{n\cdot k}\, \tilde{A}_{\rho}(k) ~.
 	\label{vertex2}
 \end{align}
The second one contains the non-abelian term $[A_{\mu},A_{\nu}]$. Exploiting the anti-symmetry
of $(S_{\mu\nu})^{\rho\sigma}$ yields
  \begin{align}
	\frac{g^2}{\lambda}\int_0^{\infty} dt\, (S_{\mu\nu})^{\rho\sigma}A_{\rho}(nt)A_{\sigma}(nt)
	=\frac{g^2}{\lambda} (S_{\mu\nu})^{\rho\sigma}
	\int \frac{d^dk}{(2\pi)^d}\int \frac{d^dl}{(2\pi)^d} \,
	\frac{1}{n\cdot (k+l)} \tilde{A}_{\rho}(k)\tilde{A}_{\sigma}(l) ~.
	\label{vertex3}
\end{align}
Neither term contains a power of $\tilde x^{\mu}$. Hence, 
the path integral can be performed precisely
as in the scalar case, thus showing that soft gluon emissions naturally exponentiate 
at NLP also for spin-one emitters.
 Finally, including the angular momomentum generator $L_{\mu\nu}$
into the exponent for gauge transformation purposes (as remarked in \cref{footnote:gauge})
 leads 
to the generalized Wilson line defined in \cref{wilson}.

\subsection{The supersymmetric model}
At this point we have achieved our goal, since
we have shown that 
 \cref{wilson} is a suitable representation for the
asymptotic states of a scattering amplitude also in the spin-one case.
However, it would be desirable to explore the connection of \cref{wilson}
with the supersymmetric worldline model, for several reasons.
The first one is that this analysis would make a better parallel with the method presented
in \cref{sec:scalar} ad \cref{dirac}. More importantly, one would like to investigate
whether there exists a wordline representation that is suitable for a propagator whose
numerator is not unity, such as for massive vector bosons or gluons in generic gauges, and more generally
for particles of higher spin. 
A complete and detailed solution to this problem is beyond the scope of this paper. 
Here,
we limit our analysis to the $N=2$ model in four dimensions\footnote{The quantization 
	in $d$-dimension is more subtle 
	(see e.g. \cite{Bastianelli:2005uy}).}, following 
the same strategy adopted in \cref{dirac} and highlighting the typical features and difficulties that one encounters
in the study of the asymptotic dynamics for spin higher than $1/2$.

We consider \cref{susy}
for
 $N=2$.
Following \cite{Howe:1989vn} and  \cite{Bastianelli:2005uy}, we first
redefine our variables via
\begin{align}
&	\psi^{\mu}=\frac{1}{\sqrt{2}}\left(\psi_1^{\mu}+i\psi_2^{\mu}\right)~,
	\qquad	\bar\psi^{\mu}=\frac{1}{\sqrt{2}}\left(\psi_1^{\mu}-i\psi_2^{\mu}\right)~,\notag\\
&	\chi^{\mu}=\frac{1}{\sqrt{2}}\left(\chi_1^{\mu}+i\chi_2^{\mu}\right)~,
	\qquad	\bar\chi^{\mu}=\frac{1}{\sqrt{2}}\left(\chi_1^{\mu}-i\chi_2^{\mu}\right)~.
	\label{complex}
\end{align}
Then, the action reads
\begin{align}
	S=~\int dt\,\left( p_{\mu}\dot{x}^{\mu}
	+ i \bar\psi^{\mu}\dot{ \psi_{\mu}}
	-H\right)~,
	\label{freeGauge}
\end{align}
where the Hamiltonian $H$ is
\begin{align}
	H&=\frac{1}{2}ep_{\mu}p^{\mu}
	+i\bar\chi\psi_{\mu}p^{\mu}
	+i\chi\bar\psi_{\mu}p^{\mu}
	-a\bar\psi_{\mu} \psi^{\mu}~.
	\label{hamiltonianGauge}
\end{align}
The tranformations for 
reparametrization invariance, $N=2$ local supersymmetry and $O(2)$ symmetry are 
respectively generated by
\begin{align}
	Q_0\equiv \frac{1}{2}p^2~, \qquad Q_1\equiv \psi\cdot p~,
	\qquad Q_2\equiv \bar \psi\cdot p~,
	\qquad J\equiv \bar \psi\cdot \psi~.
	\label{chargesGauge}
\end{align}
The $O(N)$ symmetry, with gauge field $a$,
is a distinctive feature of particles with spin $N\ge 1$. 

The quantization of this model on the closed line topology has been carried in detail 
in \cite{Bastianelli:2005vk, Bastianelli:2005uy, Bastianelli:2007pv} while
the free open propagator has been discussed in \cite{Pierri:1990rp}.
Here, in analogy with \cref{sec:scalar} and \cref{dirac}, we must consider a
path integral representation for the open line compatible with the less
common boundary conditions of asymptotic dressed propagators, i.e.
\begin{align}
	\int {\cal D} e  {\cal D} \chi {\cal D} \bar\chi {\cal D}a
	\int_{\psi (1)=\Gamma }^{\bar\psi (1)=\bar\Gamma } {\cal D}\psi {\cal D}\bar\psi
	\int_{x(0)=x_i}^{p(1)=p_f}  {\cal D}x {\cal D}p \,
	e^{ip(1)\cdot x(1)-i\int_0^1 dt\,\left(
		p\cdot \dot{x}+i\bar \psi \cdot \dot{ \psi}
		-eQ_0-\bar\chi Q_1-\chi Q_2+a J-i\epsilon
		\right)}~.
	\label{freeSpin1}
\end{align} 
Now we fix the gauge multiplet. The einbein $e$ can be set
equal to the proper time $T$, as usual. Unlike the close topology of \cite{Bastianelli:2005uy},
the Grassmann variables
$\chi$ and $\bar{\chi}$ cannot be set to zero, since they must generate
the spin structure of the propagator, in analogy with the Dirac case
of  \cref{dirac}. The gauge field $a$ deserves special attention,
since its role is to set the degrees of freedom that one wish to propagate on the worldline.
From this point of view, the choice $a=0$ is not the best one, since the corresponding
propagator carries undesired remainder terms \cite{Pierri:1990rp}. However, it 
provides a great simplification since the corresponding Faddeev-Popov determinant is trivial.
Therefore, we set
$(e,\chi,\bar{\chi},a)=(T,\theta,\bar\theta,0)$
to get
\begin{align}
	\int_0^{\infty} dT  \int d\theta d\bar\theta 
	\int_{\psi (T)=\Gamma }^{\bar\psi (T)=\bar\Gamma } {\cal D}\psi {\cal D}\bar\psi
	\int_{x(0)=x_i}^{p(T)=p_f}  {\cal D}x {\cal D}p \,
	e^{ip(T)\cdot x(T)-i\int_0^T dt\,\left(
		p\cdot \dot{x}+i \bar\psi \cdot \dot{ \psi}
		-Q_0-\frac{\bar\theta}{T} Q_1-\frac{\theta}{T} Q_2-i\epsilon
		\right)}~,
	\label{freeSpin2}
\end{align}
which can be compared with the analogous expression for the Dirac case of
 \cref{conservedDirac}.

At this point, a simple dimensional analysis reveals that 
\cref{freeSpin2} cannot yield the propagator for a vector boson field 
$A_{\mu}(x)$, since the Grassmann integrals yield
$Q_1 Q_2/Q_0$, which behaves as $\sim p^{\mu}p^{\nu}/p^2$.
In fact, the worldline representation of the $N=2$ model returns 
the propagator for the field strength tensor $F_{\mu\nu}$, rather than the fundamental field~$A_{\mu}$.
This property, which is well-known
 \cite{ Howe:1988ft, Howe:1989vn, Pierri:1990rp, Marnelius:1993ba, Bastianelli:2007pv} and shared by
all models with $N$-extended supersymmetry with $N\ge 2$ , should come as no surprise. 
It could have been guessed 
by the fact that the physical states corresponding to the quantization of the 
$N$-extended model in \cref{susy} are constructed
by taking the tensor product of the
spin variables $\psi_{\mu}^i$, which correspond to the reducible Dirac representation $(\frac{1}{2},0)\oplus(0,\frac{1}{2})$ of the Lorentz group.
However, one could argue that for our purposes this is not a huge problem, since 
the underlying idea behind the representation of a
 dressed propagator for spin $1/2$  is that what matters for the asymptotic dynamics is the denominator contribution, and not the numerator.  

In fact, the main obstacles appear when introducing a background gauge field
 \cite{Shore:1981mj, Strassler:1992zr, Reuter:1996zm}.
One might be tempted to do so by 
proceeding as we did for the spin $0$ and spin $1/2$ cases, and replace the free 
charges $Q_i$
with the corresponding $Q_i^A$, where $p_{\mu}\to p_{\mu}+g\tilde A_{\mu}$.
However, the equations of motion become inconsistent unless the field strength $\tilde F^{\mu\nu}$
of the background field is constant \cite{Buchbinder:1993ip}. Moreover,
a simple calculation reveals that the corresponding supersymmetry 
would require a vanishing~$\tilde F^{\mu\nu}$. Once again, this problem 
is shared by all models with $N\ge 2$, and is related to the consistency 
problems of theories with charged fields of spin 
higher than~$1/2$ \cite{Howe:1989vn, Gitman:1994wc}.
However, we can assume that for a soft background field the above conditions are approximately fulfilled. 
In fact, if $\tilde A_{\mu}$ is dominated by long wavelength components, the field strength is of 
order $k_{\mu}\tilde A_{\mu}$ and thus is subleading w.r.t. the hard momentum $p_f$ in the 
Lagrangian of \cref{freeSpin2}.
Hence, supersymmetry is ``softly'' broken, and 
we can repeat the previous analysis carried in \cref{dirac}.
Let us discuss this in more detail.

The fact that the background field strength vanishes in the soft limit
implies 
that 
\cref{freeSpin2} can be regarded as the expectation value
of the (approximately conserved) Noether charges $Q_1^A$ and $Q_2^A$, which can 
be evaluated at an arbitrary time.
Again, for the given boundary conditions, the proper time $T$ is a convenient choice which yields
 \begin{align}
	Q_1^A(x(T),p(T))= \Gamma \cdot (p_f-A(x(T)))~, \qquad
		Q_2^A(x(T),p(T))= \bar\Gamma \cdot (p_f-A(x(T)))~.
	\label{chargeAsymptGauge}
\end{align}
Plugging this into \cref{freeSpin2} and performing the momentum integration
around the classical solutions of \cref{eom}, we get
\begin{align}
	\langle p_f|x_i \rangle^{-1}	\langle p_f|\frac{Q_1^AQ_2^A}{2Q_0^A+i\epsilon}|x_i \rangle=
	\frac{1}{2}
	\int_0^{\infty} dT\, e^{\frac{i}{2}(p_f^2+i\epsilon)T}\int d\theta\,d\bar\theta
	e^{i (\Gamma \cdot p_f\theta+\bar \Gamma \cdot p_f\theta)}
	f(x_i,p_f,\Gamma,\bar \Gamma,T,\theta,\bar\theta)   ~,
	\label{dressedGauge2}
\end{align}
where, in analogy with \cref{ftau} and \cref{ftauDirac}, we have defined
\begin{align}
	f(x_i,p_f,\Gamma,\bar \Gamma,T,\theta,\bar\theta)&=
	\int_{\psi(T)=\Gamma, \bar\psi(T)=\bar\Gamma} {\cal D}\psi\, {\cal D}\bar\psi
	\int_{\tilde x(0)=0} {\cal D} \tilde x\, 
	e^{ -i\theta \psi\cdot A(x_i+p_f T+ \tilde{x}(T))
	-i\bar\theta \bar \psi\cdot A(x_i+p_f T+ \tilde{x}(T))
}\notag \\
	&\qquad e^{i\int_0^{T}dt\,\left(-i\bar\psi \cdot \dot{ \psi}+\frac{1}{2}\dot{ \tilde x}^2 
		+(p_f+\dot{ \tilde x})\cdot A(x_i+p_f t + \tilde x(t)) 
		+\frac{i}{2}\partial\cdot A(x_i+p_f t+ \tilde{x}(t))+\bar\psi_{\mu}\psi_{\nu}F^{\mu\nu}\right)} ~.
	\label{ftauGluon2}
\end{align}

At this point we should implement the same manipulations that we performed in 
the scalar and the Dirac cases, and truncate
the external free propagator. 
However, in this case it means that we have 
to divide by the 
free correlator $\langle F_{\mu\nu}F_{\rho\sigma} \rangle$.
Then, in analogy with
\cref{asymptoticDirac} and 
\cref{limitsDirac}, we consider the asymptotic limit $T\to\infty$
of \cref{ftauGluon2}. The effect of this limit is that the background field 
in the first line of \cref{ftauGluon2} is evaluated for asymptotic times and thus
it can be set to zero, so that the dependence of radiative factor $f(x_i,p_f,\Gamma,\bar \Gamma,T,\theta,\bar\theta)$ on $\theta$
and $\bar\theta$ can be dropped. Subsequently, the Grassmann integration over $\theta$
and $\bar\theta$
in \cref{dressedGauge2} becomes trivial and yields
 the prefactor $\Gamma\cdot p_f \bar\Gamma\cdot p_f$. 
This can be related to the numerator of the 
free correlator $\langle F_{\mu\nu}F_{\rho\sigma} \rangle$,
once a suitable representation
for the constant $\Gamma$ and $\bar\Gamma$ in terms of the 
gamma matrices is provided \cite{Howe:1989vn, Pierri:1990rp}.

To summarize, by  studying the worldline $N=2$ model with the boundary conditions 
suitably chosen to describe the asymptotic dynamics, we obtained that 
the correlator $\langle F_{\mu\nu}F_{\rho\sigma} \rangle$ in the presence of a 
an asymptotic soft background field can be expressed in terms of the radiative factor
\begin{align}
	f(x_i,p_f,\Gamma,\bar \Gamma,\infty)&=
	\int_{\psi(\infty)=\Gamma, \bar\psi(\infty)=\bar\Gamma} {\cal D}\psi\, {\cal D}\bar\psi
	\int_{\tilde x(0)=0} {\cal D} \tilde x\, 
	\notag \\
	&\qquad e^{i\int_0^{T}dt\,\left(-i\bar\psi \cdot \dot{ \psi}+\frac{1}{2}\dot{ \tilde x}^2 
		+(p_f+\dot{ \tilde x})\cdot A(x_i+p_f t + \tilde x(t)) 
		+\frac{i}{2}\partial\cdot A(x_i+p_f t+ \tilde{x}(t))+\bar\psi_{\mu}\psi_{\nu}F^{\mu\nu}
		(x_i+p_f t + \tilde x(t))
		\right)} ~.
	\label{ftauGluon2}
\end{align}
Now we can solve the path integral order by order in the soft expansion by performing
the usual rescaling $p_f^{\mu}\to \lambda n^{\mu}$ and $t\to t/\lambda$.
We start with the Grassmann integrations, and expand around 
the boundary conditions
\begin{align}
\psi^{\mu}(t)=\psi^{\mu}(T)+\chi^{\mu}(t)~,\qquad
\bar\psi^{\mu}(t)=\bar\psi^{\mu}(T)+{\bar\chi}^{\mu}(t)~.
	\label{bc}
\end{align}
Then, the path integral becomes
 \begin{align}
 	e^{\frac{i}{\lambda}\int_0^\infty dt\, \bar\Gamma_{\mu}\Gamma_{\nu}F^{\mu\nu}(x_i+p_f t + \tilde x(t))}
 	\int_{\chi(\infty)=0}^{\bar\chi(\infty)=0}
 	{\cal D}\chi{\cal D}\bar\chi\,	 e^{i\int_0^{\infty}dt\,\left(-i\bar\chi \cdot \dot{ \chi}
 		+\frac{1}{\lambda}\bar\chi_{\mu}\chi_{\nu}F^{\mu\nu}(x_i+p_f t + \tilde x(t))\right)} ~.
 	\label{ftauGluon3}
 \end{align}
Once again, the argument of $F^{\mu\nu}$ significantly simplifies by noting that $x_i^{\mu}\neq 0$
is a subleading effect, and that vertices with powers of $\tilde x^{\mu}$ and $\chi^{\mu}$ would require
additional $\tilde x$-propagators, which are suppressed in $1/\lambda$. This means that we can expand $F^{\mu\nu}$
at leading order in $\tilde x^{\mu}$, to get
 \begin{align}
	e^{\frac{i}{\lambda}\int_0^\infty dt\, \bar\Gamma_{\mu}\Gamma_{\nu}F^{\mu\nu}(n t)}
	\int_{\chi(\infty)=\bar\chi(\infty)=0} 
	{\cal D}\chi{\cal D}\bar\chi\,	 e^{i\int_0^{\infty}dt\,\left(-i\bar\chi \cdot \dot{ \chi}
		+\frac{1}{\lambda}\bar\chi_{\mu}\chi_{\nu}F^{\mu\nu}(n t)\right)} ~.
	\label{ftauGluon4}
\end{align}
In analogy with \cref{sec:den}, the path integral is Gaussian and can be reabsorbed in the 
overall normalization with the factor $\langle p_f|x_i \rangle$ of \cref{dressedGauge2}.
The prefactor, on the other hand, contains boundary information in the
term 
$[\bar\Gamma_{\mu},\Gamma_{\nu}]$. 
With a suitable representation
in terms of gamma matrices \cite{Howe:1989vn, Pierri:1990rp,  Strassler:1992zr},
this returns once again the Lorentz generator $(S_{\mu\nu})^{\rho\sigma}$, 
in agreement with \cref{vertex2} and \cref{vertex3}.

Therefore, we recover the same structure of the previous section where the GWL has been derived
without worldline fermions. Thus, 
 despite the fact that the $N=2$ model corresponds to the propagation 
of the field strength $F^{\mu\nu}$ rather than the potential $A_{\mu}$,
the denominator contribution is still given by a scalar term
and a spin dependent term that involves the Lorentz generator $(S_{\mu\nu})^{\rho\sigma}$.
This is in agreement with the analogous result obtained with a 
one-loop effective action \cite{Strassler:1992zr}, thus confirming that the 
GWL is an equivalent description where only the denominator of a dressed propagator
contributes to the asymptotics. 

It is clear that the arguments presented 
in this section can be generalized to the case of a (massive) particle of arbitrary spin,
thanks to the fact that the background field for $N>2$
is introduced in analogy with the $N=2$ case with  
 a term proportional 
 to $\psi_i^{\mu}\psi_i^{\nu}F^{\mu\nu}$  \cite{Buchbinder:1993ip, Gitman:1994wc}.
Although this is not investigated further in this work, 
the term $\psi_i^{\mu}\psi_i^{\nu}$ gives rise to 
the corresponding
Lorentz generator, once a proper representation in terms of gamma matrices is provided \cite{Howe:1988ft}.
Therefore, the derivation is similar to the spin-one case, where the numerator 
is composed of quasi-conserved and effectively free Noether charges, while 
the contribution of the background field in the denominator is coupled to the corresponding
Lorentz generator. 

%

\section{Discussion}
\label{sec:discussion}

The Generalized Wilson Line, originally proposed in \cite{Laenen:2008gt} to extend
the exponentiation of infrared radiation to next-to-leading power (NLP) and subsequently
applied in \cite{Bonocore:2016awd,Bahjat-Abbas:2019fqa} to derive factorization theorems,
is a powerful tool to describe asymptotic states dressed by soft radiation at subleading orders in 
the soft expansion. 
The derivation proposed in \cite{Laenen:2008gt} concerned essentially the case of a scalar particle dressed by next-to-soft radiation, both in the abelian and in the non-abelian case. Although 
an argument for an extension to the case of spin $1/2$ particles was presented there, some
 issue remained to be clarified, while no proof was given for spin $1$ or higher. 
 
In this work, building on the well-known supersymmetric model for a spinning
particle on the worldline, we have revisited the derivation for spin $1/2$. In particular, we have shown that the contribution of the soft background field to the numerator of an asymptotic dressed propagator vanishes. 
This fact, which was tacitly assumed in \cite{Laenen:2008gt}, is crucial to prove the truncation 
of the external free propagators in a scattering amplitude and was proven here by exploiting the supersymmetry of the corresponding worldline model. 

Then, we considered 
the spin $1$ case.
Thanks to the gauge invariance of a scattering amplitude, 
a shortcut can be used in Feynman gauge, where the numerator is unity: since no background field 
appears in the numerator, no supersymmetric model is necessary and the derivation closely follows
the scalar case. This implies that the next-to-eikonal exponentiation presented in \cite{Laenen:2008gt, Laenen:2010uz}
and the related diagrammatic analysis of webs \cite{Gardi:2010rn, Gardi:2013ita} can be naturally applied to Yang-Mills theory.   

Finally, we discussed how the GWL can be derived for particles of higher spin,
by studying the $N=2$ wordline supersymmetric model. 
The obstacle here is that this model naturally describes the propagation 
of $F_{\mu\nu}$ rather than the fundamental field $A_{\mu}$. Moreover, a general background
field is not compatible with wordline supersymmetry. However, although
the derivation of the GWL
is more challenging in this case, we presented an argument based on the observation
that the field strength for the background field vanishes in the soft limit and that 
worldline supersymmetry is only softly broken.


For both the spin $1/2$ and spin $1$ cases, the denominator contribution to the dressed propagator 
matches the one corresponding to the one-loop effective action \cite{Strassler:1992zr}, where the
first-quantized Hamiltonian is given by the squared coviariant derivative $D^2$ plus a spin dependent term $S_{\mu\nu}F^{\mu\nu}$ representing the interaction between the magnetic moment of the emitting particle and the background soft field. For a single soft emission, this result was shown in \cite{White:2014qia} to be in agreement
with the so-called tree-level next-to-soft theorems \cite{Cachazo:2014fwa}. 
Although the GWL extends this statement to all-orders in the coupling constant, 
one must be careful in applying these technique in a scattering amplitude 
with massless particles beyond the tree-level, 
since it is well-known that collinear effects are not captured by this description alone 
and must be 
compensated by radiative jets \cite{Bonocore:2016awd, Moult:2019mog, Gervais:2017yxv, Laenen:2020nrt}.

 The analogy with the
one-loop effective action is not surprising: as 
remarked in the 
introduction, at the cross-section level the external lines close at infinity, so it is natural to expect that the denominators, which do contribute to the asymptotic dynamics, must be the same
for the closed and open dressed propagators.  What we have shown in this work is that
the GWL makes this picture valid at the amplitude level. 
This corroborates the intuitive idea that asymptotic propagators are a somewhat intermediate case
between the closed and the open topologies.
In fact, although the worldine formalism has been known for some time, the quantization
of supersymmetric actions for spinning particles on the open topology is a relatively unexplored area, which might offer new insights into the structure of scattering amplitudes. The GWL offers a complementary point of view in this direction. 


The use of worldline techniques in the study of asymptotic dynamics can be also analyzed
in the light of the revived interest  in the Faddeev-Kulish coherent states 
  \cite{Kapec:2017tkm, Strominger:2017zoo, Pasterski:2017kqt, Carney:2018ygh,
 	Gonzo:2019fai, Choi:2019rlz, Pate:2019lpp , Law:2020tsg, Casali:2020vuy, Narayanan:2020amh   },
where the GWL offers a natural way to extend the analysis at subleading power in the soft expansion. 
In particular, 
the recently proposed prescription to define an infrared-finite S-matrix \cite{Hannesdottir:2019opa}
provides a modern derivation of the asymptotic Hamiltonian in terms of Wilson lines 
and effective field theory techniques. In this regard, it is noteworthy that the asymptotic Hamiltonian of the Faddeev-Kulish construction corresponds to the single-particle Hamiltonian of the worldline formalism, i.e. the inverse propagator dressed by soft radiation. Therefore, the GWL offers a nice semiclassical interpretation of the asymptotic dressed state and also a natural extension 
of this picture at subleading power that 
bypasses the long derivation of the subleading Lagrangian in the effective field theory. 
In fact, the emergence of the spin-dependence at NLP
 has a very clear origin in the GWL description and it shows
 how it affects
the asymptotics dynamics. 
Besides, 
 the quantization of the asymptotic states emerges quite neatly in this picture, since the
 Hamiltonian is derived form the standard Dirac procedure to deal with constraints.

The semiclassical description of the GWL can shed light also on
the logarithmic corrections to classical next-to-soft theorems, recently discussed 
in \cite{Sahoo:2018lxl, Laddha:2018myi}. The origin of such corrections
is due to long range forces that in four dimensions
  produce a logarithmic dependence on the proper time $T$ in the trajectory of the scattered
  particles. It is then argued that the corresponding 
  contribution to soft theorems can be obtained by replacing $\log(T)$ with $\log(k)$, where $k$
 is the soft momentum, and that such ad-hoc replacement would be presumably not needed in a quantum 
 derivation \footnote{For recent progress on this point see also \cite{A:2020lub}.}. The GWL of this paper offers a nice bridge between the classical and the quantum
 description, being
 a tool to describe radiation as perturbation of the classical path.
In particular, we have seen that the the gauge field $A_{\mu}(\tilde x(t))$
in \cref{ftau} acts as a source term on the worldline. Then, 
assuming the gauge field drops as $1/t$ in the asymptotic limit, the integral over time
in the action gives $\log(T)$, as expected. As we have described 
in \cref{sec:scalar}, this yields the effective vertices in \cref{wilson}, which in turn 
are responsible for the infrared divergences of a scattering amplitude, 
once multiple GWLs are properly inserted. 
Besides, in this work we have shown that the gauge field in the numerator
does not contribute to the asymptotic limit, and therefore we conclude that the 
logarithmic corrections mentioned above are a pure denominator effect. 

The results presented in this work can be extended in many directions. 
The most obvious one is the 
generalization of the spin-one case to gluons 
in arbitrary gauges and to massive
vector bosons. 
This problem would presumably need 
a worldline model where the spin variables 
take into account the different degrees of freedom that one wish to propagate, in 
analogy with the massive Dirac propagator that requires a fifth Grassmann variable. 
A second direction for future work is the generalization of the GWL
to soft gravitons, which has been discussed only for scalar emitters \cite{White:2011yy}.
The growing demand for precision calculations in gravitational physics, and the 
crucial role that spin effects might have to this aim \cite{Guevara:2018wpp, Vines:2016qwa, Bern:2020buy, Matas:2020wab}, make it natural to pursue this direction.  
Another aspect which is left for future work is a derivation of the GWL
where the non-abelian nature 
of the soft background field is incorporated with additional Grassmann variables, 
on the line of the results
obtained in the scalar case in \cite{Ahmadiniaz:2015xoa}. In this regard, it would 
be interesting
to investigate what is the role of such variables 
 in the soft expansion. Finally, 
there is growing evidence that the eikonal approximation
 underlies the classical limit of quantum scattering amplitudes
 \cite{Kosower:2018adc, Maybee:2019jus, delaCruz:2020bbn, Amati:1987wq, Damour:2017zjx, KoemansCollado:2019ggb}. 
  Hence,
 a comparison of the GWL for gauge bosons and gravitons 
might shed light on the worldline realization of the classical double copy \cite{Monteiro:2014cda}, which has been
 recently investigated in \cite{Goldberger:2016iau, Chester:2017vcz, Shen:2018ebu, Plefka:2019hmz, PV:2019uuv, Almeida:2020mrg}.

\section*{Acknowledgments}
The author would like to thank Fiorenzo Bastianelli, Anna Kulesza, Eric Laenen and Chris White for stimulating discussions.  

\appendix
\section{Soft expansion of the scalar asymptotic propagator}
\label{softexpansion}
We consider \cref{ftau} in the limit $T\to \infty$.
Let us start at leading power (LP) in $1/\lambda$. In this case we set $\tilde x^{\mu}=0$, which means that we ignore the fluctuations around the classical straight path 
(see \cref{eom}) and we evaluate the path integral on its stationary point. Thus
\cref{1qft} becomes
\begin{align}
	f(x_i,n\lambda,\infty)&=
	\exp\left({i\int_0^{\infty}dt\,
		n^{\mu} A_{\mu}(x_i+n t)}\right) + {\cal O}\left(\frac{1}{\lambda}\right)
	\notag \\
	&= \exp\left(-\int\frac{d^dk}{(2\pi)^d}\,\frac{n^{\mu}}{n\cdot k}
	\tilde A_{\mu}(k)\,e^{ix_i\cdot k}\right)
	+ {\cal O}\left(\frac{1}{\lambda}\right)~.
	\label{LP}
\end{align}
The term $e^{ix_i\cdot k}$ gives subleading corrections in the soft momentum $k$, so at LP we can safely set $x_i^{\mu}=0$. Therefore, 
at LP the asymptotic dressed propagator $f(x_i,n\lambda,\infty)$ reduces to the well-known
straight Wilson line.

At NLP there are two sources of corrections: the first one comes from having $x_i^{\mu}\neq 0$ in \cref{LP}. This combines with the contribution from Low's theorem\cite{Laenen:2008gt}, to give the 
orbital angular momentum $L^{\mu\nu}$. Although this contribution does not exponentiate, the separation between the orbital and the spin contributions is not gauge invariant, as observed in \cite{Bahjat-Abbas:2019fqa}. Thus, it is convenient to put also $L^{\mu\nu}$ into the exponent, albeit regarding the expression as valid up to NNLP corrections.

\begin{figure}
	\centering
	\raisebox{2.27\height}{\includegraphics[width=0.2\linewidth]{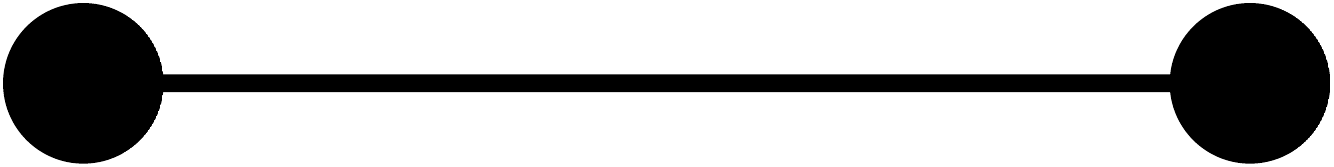}}
	\qquad \qquad \qquad
	\includegraphics[width=0.15\linewidth]{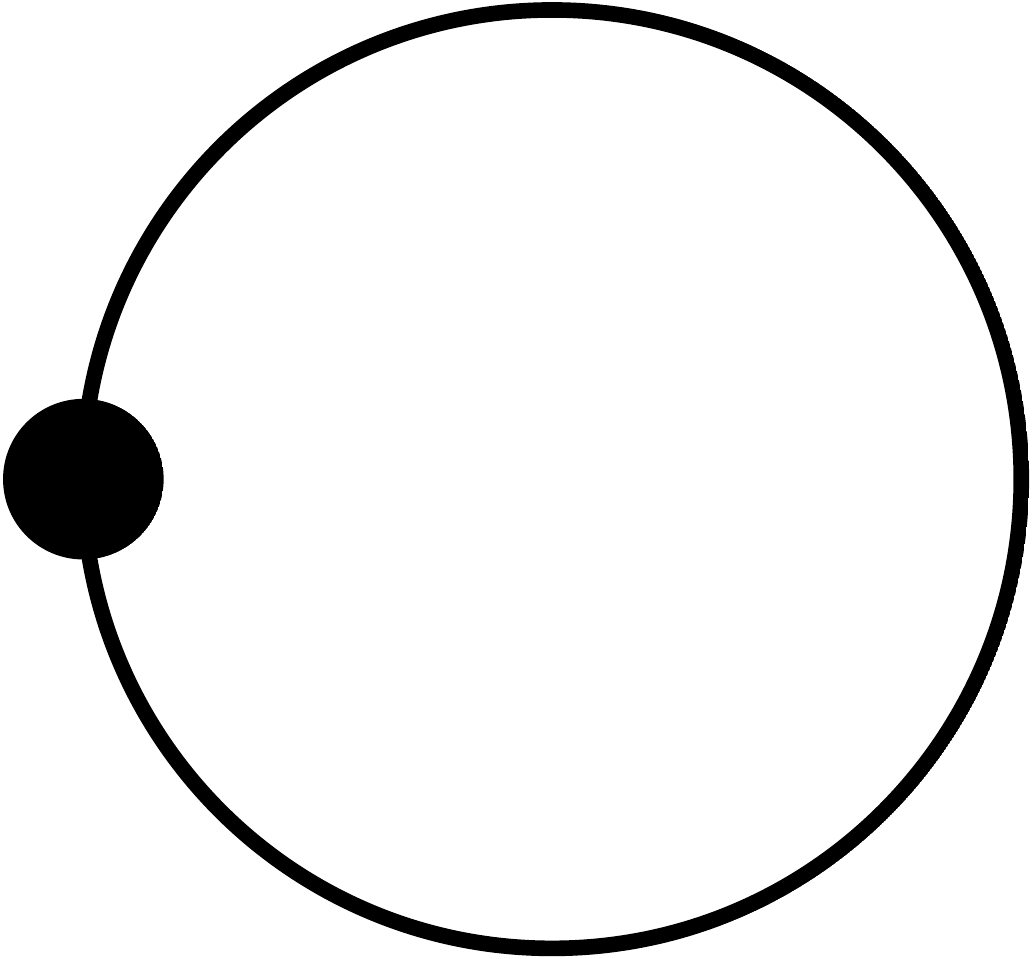}
	\caption{Connected diagrams needed up to order
		$1/\lambda$ for the evaluation of the path integral in \cref{toBeInt}.
		The diagram on the left corresponds to \cref{NLP3}, while the loop on the right
		corresponds to \cref{NLP2}.}
	\label{fig:diag}
\end{figure}

The second source of corrections comes from including quantum fluctuations in the path integral. Up to order $1/\lambda$ we need diagrams with only one propagator, which means that we need to expand the action up to second order in $\tilde x^{\mu}$ and $\dot{\tilde x}^{\mu}$. This yields
\begin{align}
	f(0,n\lambda,\infty)&=
	e^{-\frac{1}{2\lambda}\int_0^{\infty}dt\,\partial \cdot A(n t)}
	\int_{\tilde x(0)=0} {\cal D} \tilde x\, 
	\exp \Bigg(i\int_0^{\infty}dt\,\Big(\frac{\lambda}{2}\dot{ \tilde x}^2
	+\dot{\tilde x}^{\mu}A_{\mu}(n t)\notag \\
	&+(n^{\mu}+\dot{\tilde x}^{\mu})\tilde x^{\nu} \partial_{\nu}A_{\mu}(n t)
	+n^{\mu}\tilde x^{\nu}\tilde x^{\rho}
	\partial_{\nu}\partial_{\rho}A_{\mu}(n t)\Big) \Bigg)
	+ {\cal O}\left(\frac{1}{\lambda^2}\right) ~.
	\label{toBeInt}
\end{align}
The exponential with no power of $\tilde x^{\mu}$ yields 
\begin{align}   
&
	\exp \left[ \, \! \int \frac{d^d k}{(2 \pi)^d} \tilde A_\mu(k)
	\frac{k^\mu}{2\lambda n \cdot k} \right] \, 
	~.
	\label{NLP1}
\end{align}
Then, there are two class of connected diagrams with one propagator, as shown 
in \cref{fig:diag}: a 
loop-diagram with a single $\tilde x^2$ vertex, which yields
\begin{align}   
	&-
\int \frac{d^d k}{(2 \pi)^d} \tilde A_\mu(k) 
	 \frac{n^\mu k^2}{2\lambda (n \cdot k)^2} 
	~,
	\label{NLP2}
\end{align}
and a diagram where a propagator connects two $\tilde x^1$ vertices, which reads
\begin{align}   
&
	  \, \frac{1}{\lambda} \int \frac{d^d k}{(2 \pi)^d} \int \frac{d^d l}{(2 \pi)^d}
	\tilde A_\mu(k) \tilde A_\nu(l) \left(\frac{\eta^{\mu \nu}}{2 n \cdot (k + l)} - 
	\frac{n^\nu l^\mu n \cdot k + n^\mu k^\nu n \cdot l}{2 (n \cdot l)(n \cdot k) 
		\left[n \cdot (k + l) \right]} \right. \notag \\
	& \,\,\,  + \left.  \frac{(k \cdot l) n^\mu n^\nu}{2 (n \cdot l)(n \cdot k)
		\left[n \cdot (k + l) \right]} 
	\right)  \, 
	~.
		\label{NLP3}
\end{align}
It is noteworthy that \cref{NLP3} contains two gauge fields, which means that NLP soft emissions at different times are correlated pairwise along the worldline. 
More generally, at N$^n$LP we expect correlations among $n+1$ gauge bosons. 

Exponentiating the sum of the connected diagrams in \cref{NLP2} and \cref{NLP3}, and combining 
the result with \cref{LP} and \cref{NLP1}, yields the Generalized Wilson line defined in \cref{wilson} with $J^{\mu\nu}=L^{\mu\nu}$.

\bibliography{ref.bib}


\end{document}